\documentclass[aps,prd,notitlepage,showpacs,nofootinbib,superscriptaddress]{revtex4-1}
 \pdfoutput=1
\usepackage[usenames,dvipsnames]{color}  
\usepackage{graphicx}
\usepackage{setspace}
\usepackage{caption}
\usepackage{subcaption}
\usepackage{amsmath}
\usepackage{slashed}
\usepackage{amssymb}
\usepackage[colorlinks=true,citecolor=darkred,urlcolor=darkred, pdfborder={0 0 0}]{hyperref}
\usepackage[normalem]{ulem}
\usepackage{xcolor}
\usepackage{gensymb}
\usepackage{array,multirow}

\makeatletter
\def\p@subsection{}
\makeatother

\definecolor{darkred}{rgb}{0.6,0,0}

\definecolor{linkcolor}{rgb}{0,0,0.5}

\def\gsim{\raise0.3ex\hbox{$\;>$\kern-0.75em\raise-1.1ex\hbox{$\sim\;$}}}
\def\lsim{\raise0.3ex\hbox{$\;<$\kern-0.75em\raise-1.1ex\hbox{$\sim\;$}}}

\def\beqn#1{\begin{equation}\label{#1}}
\def\eeqn{\end{equation}}

\def\beqa#1{\begin{eqnarray}\label{#1}}
\def\eeqa{\end{eqnarray}}

\newcommand {\ignore}[1]{}

\def\321{$\mathrm{SU(3) \otimes SU(2) \otimes U(1)}$ }

\begin{document}

\title{Collider imprints of right handed neutrino magnetic moment operator}

\author{Eung Jin Chun}\email{ejchun@kias.re.kr}
\affiliation{Korea Institute for Advanced Study, Seoul 02455, Korea}
\author{Sanjoy Mandal}\email{smandal@kias.re.kr}
\affiliation{Korea Institute for Advanced Study, Seoul 02455, Korea}
\author{Rojalin Padhan}\email{rojalin.p@iopb.res.in}
\affiliation{Institute of Physics, Sachivalaya Marg, Bhubaneswar 751005, India}
\affiliation{Homi Bhabha National Institute, BARC Training School Complex, Anushakti Nagar, Mumbai
400094, India}

\begin{abstract}
We consider most general effective Lagrangian up to dimension five, built with Standard Model~(SM) fields and right-handed neutrinos~(RHNs) $N_i$. Assuming that the RHNs are present near the electroweak scale, we study the phenomenology of the RHNs and highlight the differences that arise due to the inclusion of dimension five operators. 
We specifically focus on the production process $e^+e^-/pp\to N_i N_j$ which comes from the dimension five magnetic moment operator. We find that this production process followed by the decay chains such as $N_i\to N_j\gamma$, $N_i\to\nu_j\gamma$ and $N_i\to\ell^\pm j j$ leads to striking collider signatures which might help to probe the Majorana nature of neutrinos. We discuss the current collider constraints on this operator, as well as projected limit at future colliders. In addition, we discuss the stellar-cooling bounds applicable to the RHN mass below 0.1 GeV.
\end{abstract}


\maketitle

\section{Introduction}
\label{sec:intro}
One of the strongest arguments in favor of new physics is the existence of neutrinos with non-zero masses~\cite{Kajita:2016cak,McDonald:2016ixn,KamLAND:2002uet,K2K:2002icj}. Explaining the origin of neutrino masses stands out as one of the biggest challenges in elementary particle physics. The simplest way to generate neutrino masses would be to add right handed neutrinos~(RHNs) to the Standard Model~(SM) particle contents and write down Yukawa interactions for neutrinos similarly to the other SM charged fermions. Then, the neutrino Yukawa coupling $Y_\nu$ has to be extremely small~$\mathcal{O}(10^{-12})$ to get the neutrino mass below 1 eV.  On the other hand, one can include the lepton-number violating Majorana mass term $M_R \overline{N_R^c} N_R$ for the RHNs which are singlet under the SM gauge group.  The smallness of the neutrino mass can then be explained through the so-called type-I seesaw mechanism: $m_\nu=Y_{\nu}^2 v^2/M_R$ \cite{Schechter:1980gr,Minkowski:1977sc,Gell-Mann:1979vob,Mohapatra:1979ia}. 
\par The coupling of neutrinos with photons might be useful to distinguish between Dirac and Majorana neutrinos as the Majorana neutrinos can only have flavor changing transition magnetic moments. However, in the SM extended with Dirac neutrinos, neutrino magnetic moments are extremely small, $\mu_\nu^{ii}\simeq 3.2\times 10^{-19}(m_i/\text{eV})\mu_B$~\cite{Giunti:2014ixa,Fujikawa:1980yx}, where $m_i$ are the light neutrino masses, $\mu_B=(e/2m_e)$ is the Bohr magneton, $e$ is the electron charge and $m_e$ is the electron mass. For Majorana neutrinos~(in the SM extended with the type-I seesaw mechanism), the transition magnetic moment is estimated to be of the order of $\sim 10^{-23}\,\mu_B$~\cite{Pal:1981rm,Giunti:2014ixa}. Clearly these values are too small to be in the reach of current experimental capabilities~\footnote{One can allow for large values of $\mu_\nu$ if one include new symmetries and interactions. Indeed, one can construct models with large magnetic moment using either extended global symmetry~\cite{Babu:1989wn,Babu:1989px} or extended gauge symmetry~\cite{Voloshin:1987qy,Barbieri:1988fh,Ecker:1989ph,Chang:1990uga,Choudhury:1990fdr}.}. Note that such a tiny magnetic moment in most of the simple extensions of SM is due to the chiral suppression and GIM cancellations. 
Contrary to this, heavy RHNs can have a large magnetic moment evading the chiral suppression and the associated phenomenologies have been discussed in Refs.~\cite{Sher:2001rk,Sher:2002ij,Aparici:2009fh,Barducci:2022gdv,Kikuchi:2008ki}. If we add only SM singlet RHNs with the Majorana mass at the electroweak scale, the simple effective field theory~(EFT) argument suggests that one can write a dimension five magnetic moment operator~(MMO) such as $\mathcal{L}=(\alpha_{NB}/\Lambda) \overline{N_R^c}\sigma_{\mu\nu}N_R B^{\mu\nu}$. If $\alpha_{NB}=\mathcal{O}(1)$ and $\Lambda=\mathcal{O}(1)$ TeV,  this gives a large magnetic moment for the RHNs. One may anticipate that the MMO will arise through a loop in realistic models, suppressing the unknown coefficient $\alpha_{NB}$.
\par Given the limited understanding on theories beyond the Standard Model (BSM),  it is useful to examine the effective Lagrangian invariant under the SM gauge symmetry in a systematic manner. The effective field theory is valid at energies below the scale of new physics denoted by $\Lambda$. In addition we will assume that the underlying physics is decoupling~\cite{Appelquist:1974tg,Weinberg:1980wa} which ensures that all the low-energy observables are suppressed by inverse powers of the cut-off scale $\Lambda$. We consider an effective Lagrangian involving SM fields and RHNs $N_i$. The EFT of this kind is denoted as $\nu$SMEFT. The use of EFTs involving RHNs  was first considered in Ref.~\cite{delAguila:2008ir} where  the complete set of operators up to dimension six has been worked out~\footnote{The Refs.~\cite{Aparici:2009fh} and \cite{Bhattacharya:2015vja} presents the non-redundant operators basis up to dimension five and dimension nine, respectively.}. Later a non-redundant basis of the effective operators was provided in Ref.~\cite{Liao:2016qyd}, and there are numerous works that cover various facets of the $\nu$SMEFT ~\cite{Mitra:2022nri,delAguila:2008ir,Aparici:2009fh,Bhattacharya:2015vja,Liao:2016qyd,Li:2021tsq,Barducci:2020icf,Aparici:2009fh,Alcaide:2019pnf,DeVries:2020jbs,Barducci:2022hll,Butterworth:2019iff}.

In this work we study the $\nu$SMEFT up to dimension five. There are three classes of operators in dimension five among which two of them contribute to the Majorana masses of the active neutrinos and RHNs, respectively, and the other  gives magnetic moments for the RHNs. We specifically focus on this last operator which generates many interesting vertices such as $Z/\gamma-N_i-N_j$ and $Z/\gamma-\nu_i-N_j$. The former coupling opens the possibility of the RHNs production through $s-$channel $Z/\gamma$ exchange at $e^+ e^-$ or $pp$ collider: $e^+e^-/pp\to Z^*/\gamma^* \to N_i N_j$. In addition to the usual RHN decay modes coming from the light-heavy neutrino mixing parameter $\theta\sim Y_\nu v/M_R$,
there will be additional decay modes such as $N_i\to N_j\gamma/N_j Z/\nu_j\gamma$ due to this MMO. From this,  many novel collider phenomenologies can arise from the production process $e^+e^-/pp\to N_i N_j$ followed by the decay chains such as $N_i\to N_j\gamma/N_j Z/\nu_j\gamma$ and $N_i\to\ell^\pm j j$. Two of the most interesting collider signatures we find are $\ell_a^\pm \ell_b^\pm + 4j$ and $\ell_a^\pm \ell_b^\pm + 4j+\gamma$ which clearly violates lepton number by two unit and might be relevant to probe the Majorana nature of neutrinos~\footnote{Note that the same final state $\ell_a^\pm \ell_b^\pm + 4j$ is also present in usual type-I seesaw model but are very much suppressed as the production process $e^+e^-/pp\to N_i N_j$ is proportional to $|\theta|^4$, where $\theta$ is the light-heavy neutrino mixing.}.
\par This article is organised as follows. In Sec.~\ref{sec:model}, we begin with the general set up and construct dimension five operators including MMO and discuss their physical aspects. In Sec.~\ref{sec:production}, we discuss the production cross-section for heavy neutrinos through MMO at $e^+e^-$ and $pp$ collider. Further in Sec.~\ref{sec:decay}, we discuss the various RHN decay modes and their decay lengths in presence of both the light-heavy neutrino mixing and MMO. Following the discussion of the RHN production and their decay modes, in Sec.~\ref{sec:collider-signatures} we presented possible interesting collider signatures. In Sec.~\ref{sec:constraints}, we discuss constraint on MMO coming from the existing collider searches and investigate the sensitivity on MMO at the future colliders such as ILC, FCC-ee and Muon colliders. In Sec.~\ref{sec:astro}, we review the constraints on MMO coming from cosmology and astrophysical observables. Finally in Sec.~\ref{sec:conclusion}, we summarise our main findings.
\section{Model set up}
\label{sec:model}
We will assume that there are two SM singlet RHNs for simplicity and the most general form of the effective Lagrangian up to dimension five including these RHNs:
\begin{align}
\mathcal{L}=\mathcal{L}_{\rm SM}+i\overline{N}_R\partial N_R - \overline{L}_{L}Y_\nu \tilde{H} N_R - \frac{1}{2} M_R \overline{N_R^c} N_R + \frac{\mathcal{O}^5}{\Lambda}+\text{H.c},
\label{eq:Lag}
\end{align}
where $\tilde{H} = i\sigma^{2}H^{*}$, $N^{c}_{R}$ = $C\bar{N}^{T}_{R}$ with charge conjugation matrix $C = i\gamma^{2}\gamma^{0}$, $M_{R}$ stands for the Majorana mass of a RHN, $L_{L}$ is the SM lepton doublet and $Y_\nu$ is the Dirac-type Yukawa coupling. The $\mathcal{O}^5$ are dimension five operators suppressed by the cut-off scale $\Lambda$. There are the following three types of dimension five effective operators:
\begin{itemize}
\item $\mathcal{O}^5_W=\alpha_W (\overline{L_L^c}\tilde{H})(\tilde{H}^\dagger L_L)$: This is known as Weinberg operator and the couplings $\alpha_W$ is symmetric in the flavour indices due to the Lorenz invariance. The Weinberg operator $\mathcal{O}^5_W$ only involves SM fields which primarily contributes to active neutrino masses.
\item $\mathcal{O}^5_{NH}=\alpha_{NH}\overline{N_R^c} N_R (H^\dagger H)$: The coupling $\alpha_{NH}$ is symmetric in the flavour indices due to the Lorenz invariance. The operator $\mathcal{O}^5_{NH}$ provides additional contribution to the RHN Majorana mass which we will discuss later. This operator also gives additional neutrino-Higgs interactions and can give new decay channel such as $h\to NN$~if allowed kinematically.
\item $\mathcal{O}^5_{NB}=\alpha_{NB}\overline{N_R^c}\sigma_{\mu\nu}N_R B^{\mu\nu}$, where $\sigma_{\mu\nu}=\frac{i}{2}[\gamma_{\mu},\gamma_{\nu}]$ and $B_{\mu\nu}=\partial_{\mu}B_{\nu}-\partial_{\nu}B_{\mu}$ is the field strength tensor corresponds to $U(1)_Y$ gauge group. The coefficient $\alpha_{NB}$ is an antisymmetric matrix which arises if we only consider more than one RHN. Although, this operator does not play any role in the neutrino mass matrix but provides non trivial vertices between heavy neutrinos and SM neutral vector bosons $Z/\gamma$.
\end{itemize}
Out of these three operators, $\mathcal{O}^5_W$ and $\mathcal{O}^5_{NH}$ may be formed at the tree level, while $\mathcal{O}^5_{NB}$ would only exist via loop mediated processes, assuming the full theory is a gauge theory~\footnote{Possible example of UV completions include models with additional scalar and fermions or models with additional vectors and fermions, with non-vanishing hypercharge~\cite{Aparici:2009fh,Aparici:2009oua}.}. As a consequence, one can estimate a further $\frac{1}{16\pi^2}$ suppression to the $\alpha_{NB}$ coefficient~\cite{Arzt:1994gp}. These type of UV completion are generally referred to as weakly coupled~\cite{Buchmuller:1985jz,Craig:2019wmo}. An alternative possibility could be for the MMO to be generated by some strong dynamics~\cite{Manohar:1983md,Georgi:1992dw}. In this case, one can adopt the convenient parametrization of the MMO as $\alpha_{NB}/\Lambda\,\, \overline{N_R^c}\sigma_{\mu\nu}N_R B^{\mu\nu}$ with $\alpha_{NB}=\mathcal{O}(1)$. In our analysis, we identify the scale associated with MMO as $\Lambda$, assuming coupling $\alpha_{NB}$ to be order one~\footnote{For weakly coupled case, one should interpret the actual cut-off scale as $\Lambda/(16\pi^2)$.}.\\
{\bf Neutrino mass in dimension five:} We now define the neutrino mass matrix while considering all the relevant terms up to dimension five. After the electroweak symmetry breaking the Lagrangian $\mathcal{L}$ in Eq.~\ref{eq:Lag} will give the following neutrino mass matrix in the basis $(\nu_L \,\,  N_R^c)$,
\begin{align}
\mathcal{L}_m=-\frac{1}{2}\begin{pmatrix} \overline{\nu}_L &  \overline{N_R^c} \end{pmatrix}  \begin{pmatrix} -\frac{\alpha_W}{\Lambda} v^2   &    m_D\\
m_D^T    &    \tilde{M}_R  \end {pmatrix}   \begin{pmatrix}    \nu_L^c \\
N_R\end{pmatrix} + \text{H.c} \text{    with      }m_D=\frac{Y_\nu v}{\sqrt{2}}\,\,\text{and}\,\, \tilde{M}_R=M_R - \frac{\alpha_{NH}}{\Lambda} v^2,
\label{eq:Lm}
\end{align}
where $v$ is the SM Higgs vacuum expectation value~(VEV). For the sake of simplicity, we consider only two RHNs. It is straightforward to extend with more RHNs. Having two RHNs, one of the active neutrinos will be left massless. The full $5\times 5$ neutrino mass matrix in Eq.~\ref{eq:Lm} can then be diagonalized by a unitary matrix $\mathcal{U}^\dagger \mathcal{M}_\nu \mathcal{U}^*=\mathcal{M}_\nu^{\rm diag}$, where $\mathcal{U}$ is a product of a block-diagonalization followed by separate diagonalizations in the light and heavy sectors. Approximately this can be expressed as
\begin{align}\label{eq:u-bdiag-1}
\mathcal{U} & \approx
\left(
\begin{array}{cc}
1-\frac{1}{2}\theta\theta^{\dagger} & \theta \\
-\theta^{\dagger} & 1-\frac{1}{2}\theta^\dagger\theta
\end{array} \right) \left(
\begin{array}{cc}
U_{\rm lep} & 0\\
0  &   V_R
\end{array} \right),
\end{align}
where $\theta^*=m_D \tilde{M}_R^{-1}$ is the mixing angle between the light and heavy neutrinos and $U_{\rm lep}$ is the leptonic mixing matrix. Without loss of generality one can assume $M_R$ to be diagonal. Furthermore, if $\alpha_{NH}$ is diagonal then $\tilde{M}_R$ will be also diagonal and hence one can take $V_R=\mathit{I}_2$. Note that we take the usual seesaw approximation $\tilde{M}_R\gg m_D$ (that is, $|\theta|\ll 1$) leads to the following effective light and heavy neutrino Majorana mass matrix,
\begin{align}
m_\nu\approx -\frac{\alpha_W}{\Lambda}v^2 - m_D \tilde{M}_R^{-1} m_D^T\,\,\,  \text{  and  } \,\,\, M_N\approx \tilde{M}_R.
\end{align} 
{\bf Charged, neutral and Yukawa interactions:} The flavor state $\nu_L$ and $N_R$ are related with the mass eigenstate $\nu_m$ and $N_{m'}$ as follows,
\begin{align}
&\nu_L=P_L \begin{pmatrix} K_L &  K_H \end{pmatrix} \begin{pmatrix}  \nu_m \\
N_{m'}
\end{pmatrix}  \text{   with   } K_L=(1-\frac{1}{2}\theta\theta^\dagger) U_{\rm lep},  \,\, K_H=\theta,\\
& N_R=P_R\begin{pmatrix} -\theta^T &  V_R \end{pmatrix}  \begin{pmatrix}  \nu_m \\
N_{m'}
\end{pmatrix},
\label{eq:mixing-relation}
\end{align}
where $P_{L,R}=\frac{1}{2}(1\mp\gamma_5)$. Due to the mixing between light and heavy neutrinos, charged and neutral current interaction will be modified. The charged current interaction takes the following form,
\begin{align}
-\mathcal{L}_{\rm CC}=\frac{g}{\sqrt{2}}\bar{\ell} \gamma_\mu P_L K n W^\mu + \text{H.c.,}  = \frac{g}{\sqrt{2}} \left(\bar{\ell} \gamma_\mu P_L K_L\nu + \bar{\ell} \gamma_\mu P_L K_H N\right) W^\mu + \text{H.c.},
\label{eq:CC}
\end{align}
where $n=(\nu \,\, N)$ and $K=(K_L\,\, K_H)$. The neutral current and Yukawa interaction takes the following form,
\begin{align}
-\mathcal{L}_{\rm NC}& =\frac{g}{2\sin\theta_W} \bar{n}\gamma_\mu P_L (K^\dagger K) n Z^\mu, \nonumber \\
& \approx \frac{g}{2\sin\theta_W}\left(\bar{\nu}\gamma_\mu P_L (K_L^\dagger K_L) \nu + \bar{\nu}\gamma_\mu P_L (K_L^\dagger K_H) N + \bar{N}\gamma_\mu P_L (K_H^\dagger K_L) \nu \right) Z^\mu,\\
-\mathcal{L}_{\rm Yukawa}& = \frac{(v+h)}{\sqrt{2}}\,\overline{\nu_L} Y_\nu P_R N + \text{H.c}.
\label{eq:NCandYuk}
\end{align}
Here we have neglected the terms  proportional to $\mathcal{O}(\theta^2)$. Now let us discuss the relevant contribution coming from dimension five operators after the electroweak symmetry breaking. With the definition of $N_R$ as given in Eq.~\ref{eq:mixing-relation}, dimension five operator $\mathcal{O}^5_{NH}$ can be expanded as
\begin{align}
-\mathcal{L}_{NH} & =\frac{(v+h)^2}{2\Lambda}\Big\{\overline{N}\left(\alpha_{NH}P_R+\alpha_{NH}^{\dagger}P_L\right) N-\overline{\nu}\left(\theta\alpha_{NH}P_R+\theta^{*}\alpha_{NH}^{\dagger}P_L\right) N \nonumber\\
&- \overline{N}\left(\alpha_{NH}\theta^{T}P_R+\alpha_{NH}^{\dagger}\theta^{\dagger}P_L\right)\nu\Big\} .
\end{align}
Hence, this operator gives interaction vertices such as $h-h-N-N$ and $h-N-N$. Other vertices such as $h-h-\nu-N$ and $h-\nu-N$ are  suppressed by the mixing angle $\theta$. Similarly, substituting Eq.~\ref{eq:mixing-relation}, dimension five operator $\mathcal{O}^5_{NB}$ can be expanded as
\begin{align}
-\mathcal{L}_{NB} & =\frac{1}{\Lambda}\Big\{ \overline{N}\sigma_{\mu\nu}\left(\alpha_{NB}P_R + \alpha_{NB}^{\dagger} P_L\right) N - \overline{\nu}\sigma_{\mu\nu}\left(\theta\alpha_{NB}P_R + \theta^{*}\alpha_{NB}^{\dagger}P_L\right) N \nonumber\\
 & - \overline{N}\sigma_{\mu\nu}\left(\alpha_{NB}\theta^T P_R + \alpha_{NB}^{\dagger}\theta^{\dagger} P_L\right) \nu\Big\} \left(c_W F_{\mu\nu}- s_W Z_{\mu\nu} \right),
 \label{eq:LNB}
\end{align}
where $c_W=\cos\theta_W$ and $s_W=\sin\theta_W$, with $\theta_W$ the weak mixing angle. $F_{\mu\nu}$ and $Z_{\mu\nu}$ are the Abelian field strengths of the photon and the $Z$ boson. We see that the electroweak moment operator $\mathcal{O}^5_{NB}$ generates many interesting vertices including $Z/\gamma-N_i-N_j$~(with $i\neq j$) and $Z/\gamma-\nu_i-N_j$. The latter coupling is mixing suppressed.
\\
{\bf Estimation of mixing parameter $\theta$:} With two RHNs and neglecting the contribution coming from operator $\mathcal{O}^5_W$~\footnote{We assume that the seesaw contribution to the active neutrino masses predominates over the ones caused by the effective operators, as the validity of the $\nu$SMEFT requires $v,\, M_N\ll\Lambda$.}, we can make use of an adapted Casas-Ibara parametrisation~\cite{Casas:2001sr,Ibarra:2003up} to obtain an expression for the Yukawa matrix $Y_\nu$:
\begin{align}
Y_\nu\approx \frac{\sqrt{2}}{v}U_{\rm lep}^\dagger\sqrt{m_{\rm light}}R\sqrt{M_N},
\end{align}
where $M_{N}=\text{diag}(M_{N_1},M_{N_2})$, $U_{\text{lep}}$ is the leptonic mixing matrix diagonalizing the neutrino mass matrix: $U_{\text{lep}}^{T} \, m_\nu \, U_{\text{lep}}=m_{\rm light}$ with $m_{\rm light}=\text{Diag}(m_1,m_2,m_3)$. Note that $R$ is a $3\times 2$ complex orthogonal matrix which can be written as 
\begin{align}
R=\begin{pmatrix}
0  &  0\\
\cos y &  \sin y \\
-\sin y &  \cos y
\end{pmatrix}  \text{  for  {\bf N}O}~(m_1=0)\text{  and  }
R=\begin{pmatrix}
\cos y &  \sin y \\
-\sin y &  \cos y \\
0  &  0
\end{pmatrix}  \text{  for  {\bf IO}}~(m_3=0).
\end{align}
where {\bf NO}~({\bf IO}) stand for normal~(inverted) ordering. We will use the standard angular parametrization of the leptonic mixing matrix:
\begin{eqnarray}
U_{\rm lep} = \left(\begin{array}{lll}  c_{12} c_{13} & s_{12} c_{13} & s_{13} e^{-i\delta} \\
-s_{12} c_{23} - c_{12} s_{23} s_{13} e^{i\delta}  & c_{12} c_{23} - s_{12} s_{23} s_{13} e^{i \delta} & s_{23} c_{13}  \\
s_{12} s_{23} - c_{12} c_{23} s_{13} e^{i \delta} & -c_{12} s_{23} - s_{12} c_{23} s_{13} e^{i\delta} & c_{23} c_{13} \end{array}\right) U_{\rm ph} ,
\label{eq:pmns1}
\end{eqnarray}
\begin{figure}[!t]	
	\centering
	\includegraphics[width=0.4\linewidth]{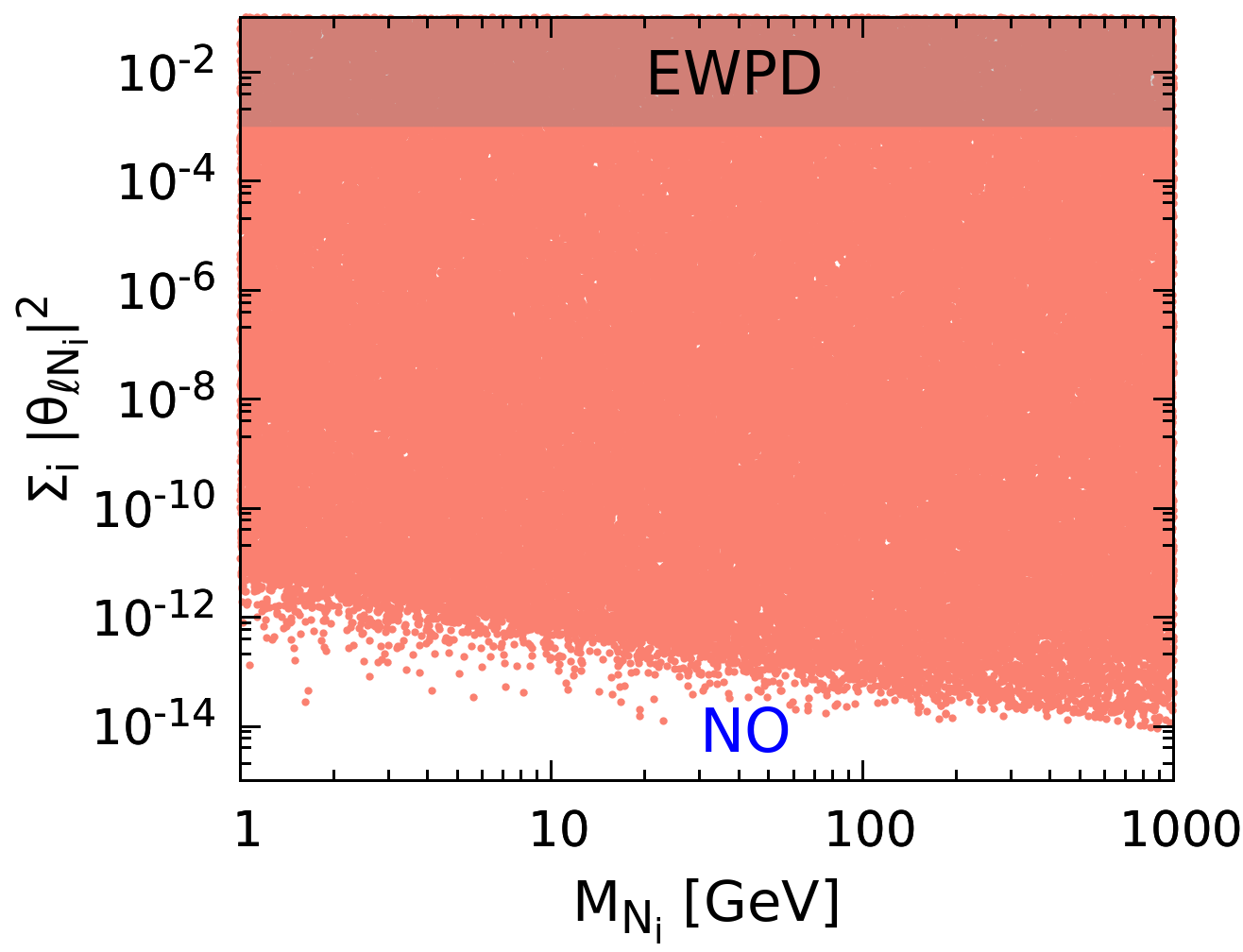}~~~~
	\includegraphics[width=0.4\linewidth]{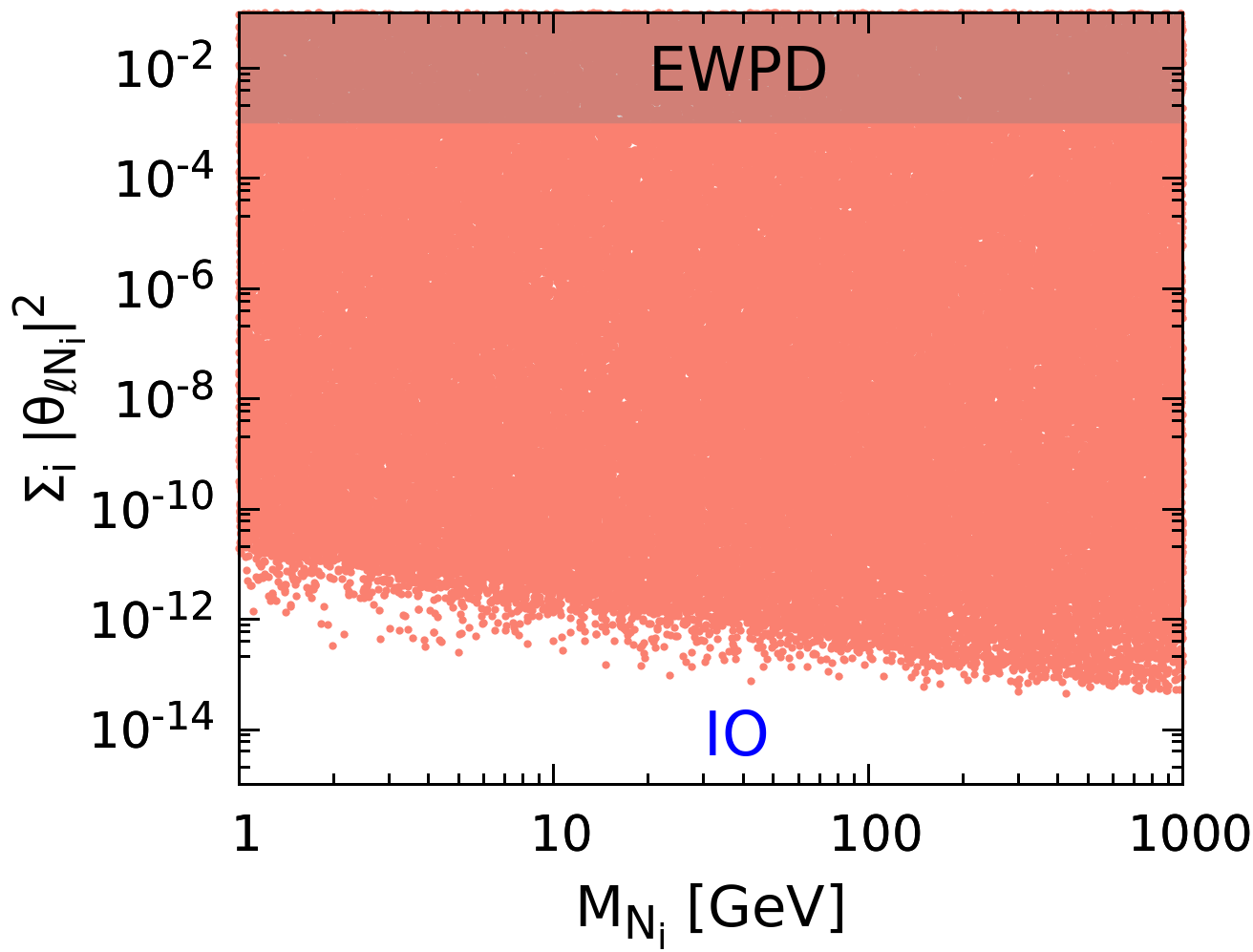}
	\caption{Variation of light-heavy neutrino mixing $\sum_{i}|\theta_{\ell N_i}|^2$ as a function of heavy neutrino mass $M_{N_i}$ in the {\bf NO}~(left panel) and {\bf IO}~(right panel) case where $\ell$ stands for either $e$, $\mu$ or $\tau$. For this we fixed the oscillation parameters to their best fit values~\cite{deSalas:2020pgw} and varied the other relevant parameters as follows: $0\leq \delta_{\rm CP}\leq 2\pi$, $1\leq M_{N_i}\leq 1000$~GeV, $0\leq a\leq 2\pi$ and $0\leq b \leq 12$. For simplicity we consider Majorana phase to be zero. See text for more details. }
	\label{fig:Mixing}
\end{figure}
where $U_{\rm ph}$ contains the Majorana phases and can be parameterized as $U_{\rm ph}=\text{Diag}(e^{-i\alpha}, e^{i\alpha}, 1)$. To check the impact of $R$ matrix on the prediction of the mixing $\theta$, let us write the complex angle $y$ as $y=a+i b$ and define an order parameter as $\theta_p^2=\text{Tr}(\theta^\dagger\theta$). The order parameter $\theta_p^2$ then can be expressed in terms of neutrino mass parameters and $a,b$ as
\small
\begin{align}
& \theta_{p,\textbf{NO}}^2=-\frac{(m_3-m_2)\delta M_N \cos(2a)}{2 M_{N_1} M_{N_2}}+\frac{(m_2+m_3)(M_{N_1}+M_{N_2})\cosh(2b)}{2 M_{N_1} M_{N_2}}\xrightarrow{\text{large b}} \frac{(m_2+m_3)(M_{N_1}+M_{N_2})\text{exp}(2b)}{2 M_{N_1} M_{N_2}},\\
& \theta_{p,\textbf{IO}}^2=-\frac{(m_2-m_1)\delta M_N \cos(2a)}{2 M_{N_1} M_{N_2}}+\frac{(m_1+m_2)(M_{N_1}+M_{N_2})\cosh(2b)}{2 M_{N_1} M_{N_2}}\xrightarrow{\text{large b}} \frac{(m_1+m_2)(M_{N_1}+M_{N_2})\text{exp}(2b)}{2 M_{N_1} M_{N_2}},
\label{eq:thetap}
\end{align}
\normalsize
where  $\delta M_N=(M_{N_2}-M_{N_1})$. Note that with the limit $a,b\to 0$ one recovers the naive seesaw relation $\theta_p^2=m_{\rm light}/M_N$. In Eq.~\ref{eq:thetap}, the first term will be always subdominant and we clearly see that if the imaginary value of $b$  for the complex orthogonal matrix is large, the mixing angles can be hugely enhanced. An upper limit on the values of $b$ is set by the electroweak precision data~(EWPD) requiring $|\theta|^2<\mathcal{O}(10^{-3})$~\cite{Antusch:2014woa}. In Fig.~\ref{fig:Mixing} we show how mixing parameter $\theta_{\ell N_i}$ varies with heavy neutrino mass $M_{N_i}$ for both the {\bf NO}~(left panel) and {\bf IO}~(right panel). We have varied the parameters $a,b$ in the range $0\leq a\leq 2\pi$ and $0\leq b \leq b_{\rm max}$ where we find that one should take approximately $b_{\rm max}<12$ to be consistent with EWPD constraints. We find the overall behaviour of the mixing parameter $\sum_i|\theta_{\ell N_i}|^2$ is same for $\ell=e,\mu$ and $\tau$. Note that the lower limit on mixing parameter is $|\theta_{\ell N_i}|^2=10^{-14}$  coming from the naive seesaw relation $|\theta_{\ell N_i}|^2\sim m_{\rm light}/M_N$ and the gray region is excluded from EWPD-data~\footnote{Note that there exists various type of constraints~\cite{deGouvea:2015euy,Chrzaszcz:2019inj,ATLAS:2019kpx,CMS:2018iaf,ATLAS:2015gtp,CMS:2018jxx,Das:2017zjc,Dercks:2018wum,L3:1992xaz,L3:2001zfe,Banerjee:2015gca,Das:2023tna,Chun:2019nwi,Das:2018usr} on the mass-mixing plane which we do not show here to avoid overcrowding the plot.}. The upshot of the above discussion is that for all practical purposes the masses and mixings can be treated as free parameters and constrained only by experimental observations.
\section{Production of heavy neutrinos at lepton and hadron colliders}
\label{sec:production}
The heavy RHN neutrinos can be produced at $e^+e^-$ or $pp$ collider from a variety of production modes. At $pp$ colliders, the most extensively studied collider production mechanism is the charged current and neutral current Drell-Yan process, $pp\to W^{\pm *}\to\ell^{\pm} N$ and $pp\to Z^{*}\to \nu N$. At $e^+e^-$ collider, the heavy neutrino can be produced as $e^+e^-\to\nu N$ through $W$ and $Z$ mediated t and s-channel processes, respectively. However, barring resonant production, the heavy neutrinos production cross-section is very small due to light-heavy neutrino mixing~($\mathcal{O}(m_D\tilde{M}_R^{-1})$) suppression.
\begin{figure}[!htbp]
	\centering
	\includegraphics[width=0.45\linewidth]{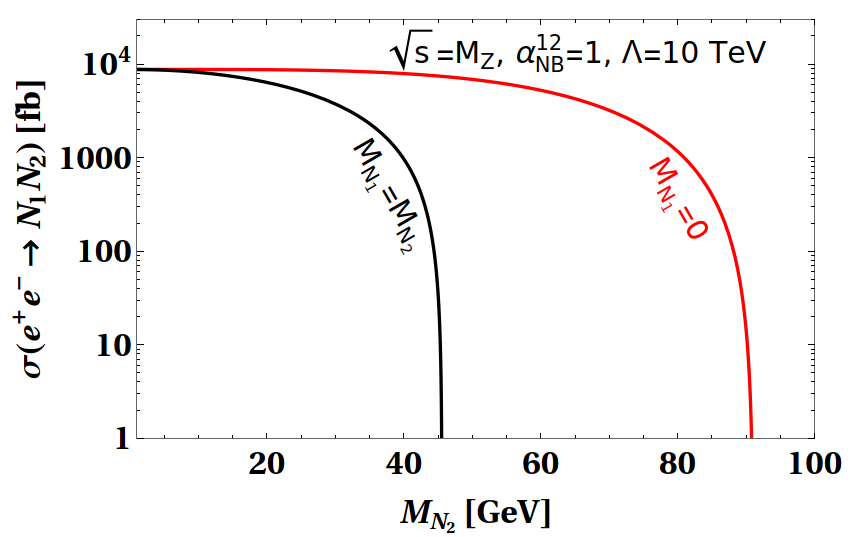}
	\includegraphics[width=0.45\linewidth]{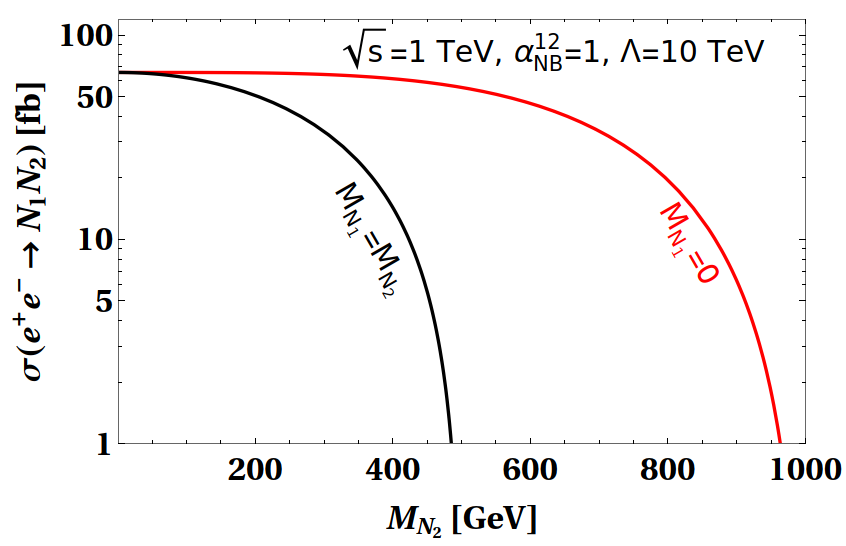}
	\includegraphics[width=0.45\linewidth]{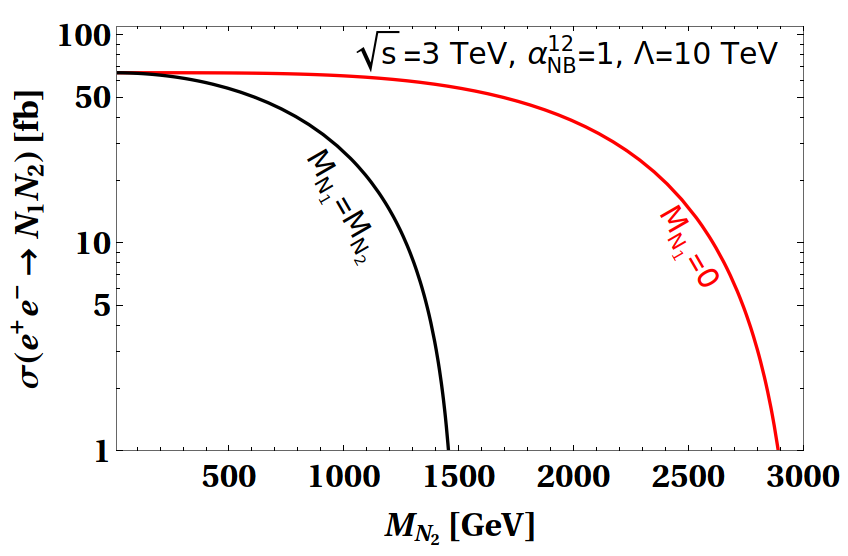}
	\caption{$e^+e^-\to N_1 N_2$ cross-section as a function of the heavy neutrino mass, $M_{N_2}$, for different center-of-mass energies. The black and red lines stand for the case of $M_{N_1}=M_{N_2}$ and $M_{N_1}=0$, respectively. We took $\alpha_{NB}^{12}=1$ and $\Lambda=10$~TeV.}
	\label{fig:eeXS}
\end{figure}
\begin{figure}[!htbp]
	\centering
	\includegraphics[width=0.45\linewidth]{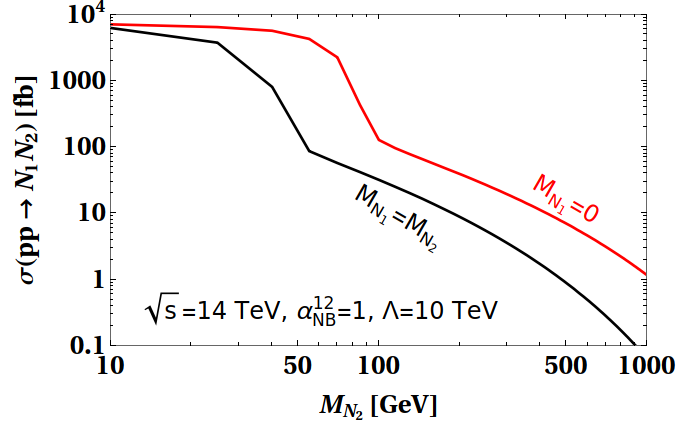}
		\includegraphics[width=0.45\linewidth]{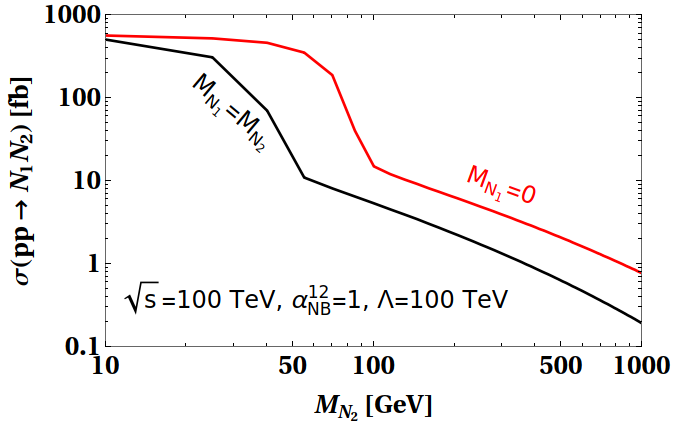}
	\caption{ $pp\to N_1 N_2$ cross-section at $pp$ collider as a function of the mass $M_{N_2}$. The left and right panel is for center-of-mass energy $\sqrt{s}=14$~TeV and $\sqrt{s}=100$~TeV, respectively. The red and black line stands for the assumption $M_{N_1}=0$ and $M_{N_1}=M_{N_2}$, respectively. We fix $\alpha_{NB}^{12}=1$ and took $\Lambda=10$~TeV~(100 TeV) for $\sqrt{s}=14$ TeV~(100 TeV) to ensure the validity of effective theory, $\sqrt{\hat{s}}\ll \Lambda$.}
	\label{fig:ppXS}
\end{figure}
Apart from the heavy neutrino production through light-heavy neutrino mixing, in presence of dimension five operator $\mathcal{O}^5_{NB}$, one has the possibility of heavy neutrino production through s-channel $\gamma/Z$ exchange at $e^+e^-$ or $pp$ collider as $pp,e^+e^-\to N_1 N_2$. In Fig.~\ref{fig:eeXS} and \ref{fig:ppXS}, we show the production cross-section for $N_1 N_2$ at $e^+ e^-$ and $pp$ collider, respectively. For $e^+ e^-$ collider we show the cross-section for center-of-mass energies $\sqrt{s}=M_Z$, 1 TeV and 3 TeV, whereas for $pp$ collider we consider two center-of-mass energies $\sqrt{s}=14$~TeV and 100 TeV. In each panel the black and red lines stand for $M_{N_1}=M_{N_2}$ and $M_{N_1}=0$, respectively. We see that as long as the magnetic moment interaction is strong enough the heavy neutrino productions can be large and might be dominant over the production through the light-heavy neutrino mixing.
\par We see that, for the lepton collider, cross-sections are almost independent of the center-of-mass energy (except for the $Z$-resonance) and are of the order of $\mathcal{O}(100\,\text{fb})$.
Near the $Z$-resonance, the cross-section is enhanced and can be of order of $\mathcal{O}(10^4\, \text{fb})$. On the other hand, for $pp$ collider, the cross-section can be large only near the $Z$-resonance, i.e. only when $M_{N_1}+M_{N_2}< M_Z$. The cross-section decreases rapidly for larger masses. 
\section{Heavy neutrino decay modes}
\label{sec:decay}
Before discussing the impact of the new dimension five effective operator on the phenomenology of RHNs at colliders, let us first examine the dominant decays of the singlet fermions $N_i$ considering the mass hierarchy $M_{N_2}\geq M_{N_1}$.
Depending on the masses, the heavy neutrinos will decay either via two body or three body modes. If we only take into account the renormalisable operators, heavy neutrinos can decay to $\ell W^{(*)}$, $\nu Z^{(*)}$ and $\nu h^{(*)}$ via the light-heavy neutrino mixing. Including the dimension five operators, there appear four additional decay modes: $N_{i}\to N_{j}\gamma/N_j Z~(i\neq j)$, $N_i\to\nu_j\gamma$ and $N_i\to N_j h$ coming from the operators $\mathcal{O}^5_{NB}$ and $\mathcal{O}^5_{NH}$, respectively. These operators also contribute to the decay modes $N_{i}\to \nu_j Z$ and $N_i\to\nu_j h$, but these contributions are proportional to $(\theta\alpha_{NH/NB})^2$, and thus suppressed compare to the contribution from the renormalisable terms. 
In Table.~\ref{tab:Decay}, we summarized all possible decay modes as well as the operators responsible for these decay modes. The explicit form of the all possible decay channels are listed in Appendix.~\ref{app:decay-widths}.
\begin{table}[h!]
\centering
\begin{tabular}{|c||c|c|}
\hline
 Decay modes &  Contributing Operators  &  Contributions \\
\hline
$ N_{i} \to \ell_{j} W$ &  $\mathcal{L}_{\rm CC}$ &  $\theta^2$\\
\hline
$ N_{i} \to \nu_{j} Z $  & $\mathcal{L}_{\rm NC}$, $\mathcal{O}^5_{NB}$  & $\theta^2$, $(\theta\alpha_{NB})^2$\\
\hline
$ N_{i} \to \nu_{j} h $  & $\mathcal{L}_{\rm Yukawa}$, $\mathcal{O}^5_{NH}$  &  $\theta^2$, $(\theta\alpha_{NH})^2$ \\
\hline
$ N_{i} \to \nu_{j} \gamma $ & $\mathcal{O}^5_{NB}$ & $(\theta\alpha_{NB})^2$  \\
\hline
$ N_{i} \to N_{j} \gamma $; $i\neq j$ & $\mathcal{O}^5_{NB}$ & $ \alpha_{NB}^2 $  \\
\hline
$ N_{i} \to N_{j} h $ & $\mathcal{O}^5_{NH}$, $\mathcal{L}_{\rm Yukawa}$ & $ \alpha_{NH}^2 $, $\theta^4$  \\
\hline
$ N_{i} \to N_{j} Z $; $i\neq j$ & $\mathcal{O}^5_{NB}$, $\mathcal{L}_{\rm NC}$ & $ \alpha_{NB}^2 $, $\theta^4$  \\
\hline
\end{tabular}
\caption{Different possible two body decay modes along with the operators that can contribute to these decays.}
\label{tab:Decay}
\end{table}
\begin{figure}[!htbp]
	\centering
	\includegraphics[width=0.45\linewidth]{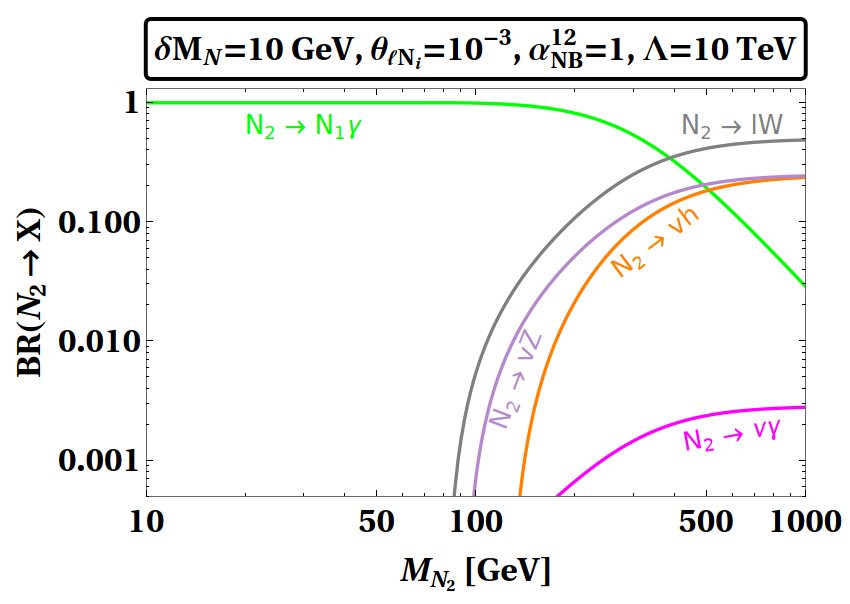}
	\includegraphics[width=0.45\linewidth]{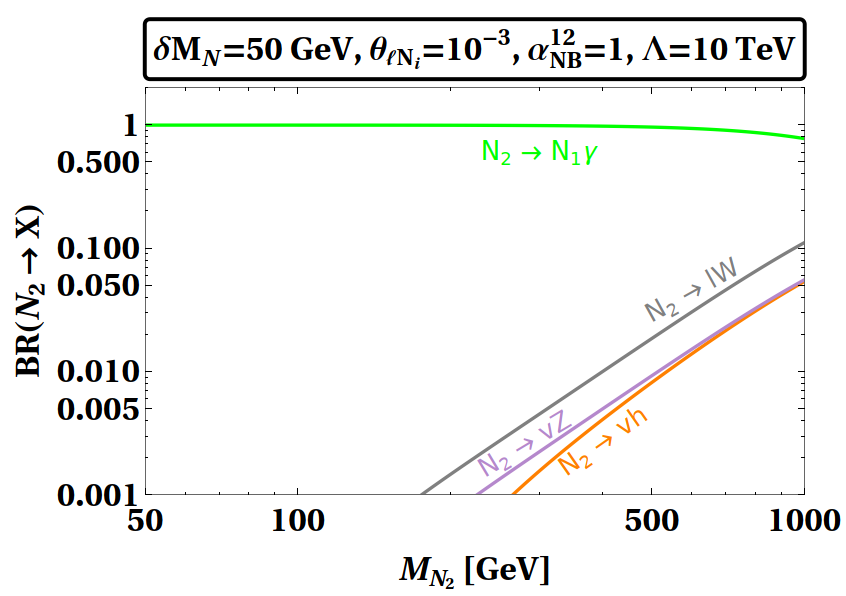}
        \includegraphics[width=0.45\linewidth]{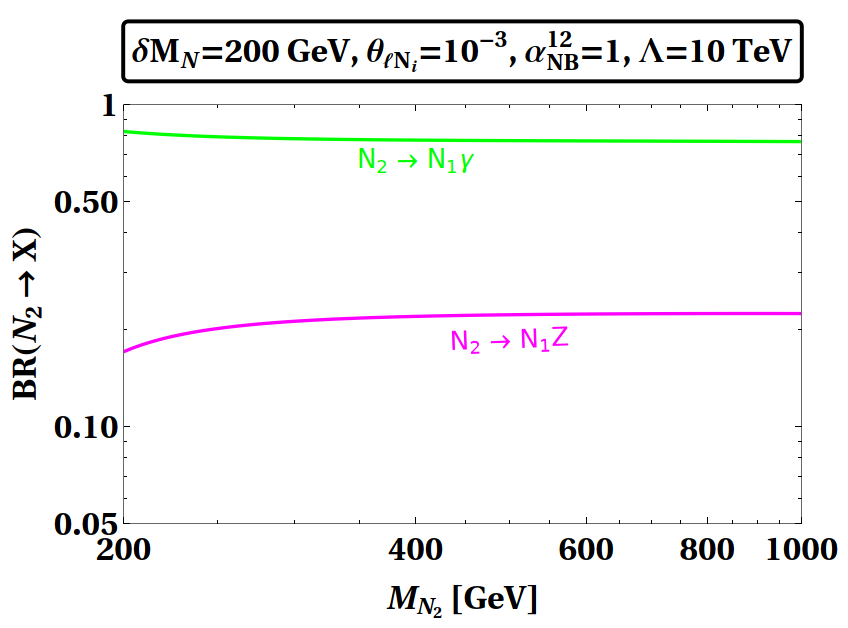}
        \includegraphics[width=0.45\linewidth]{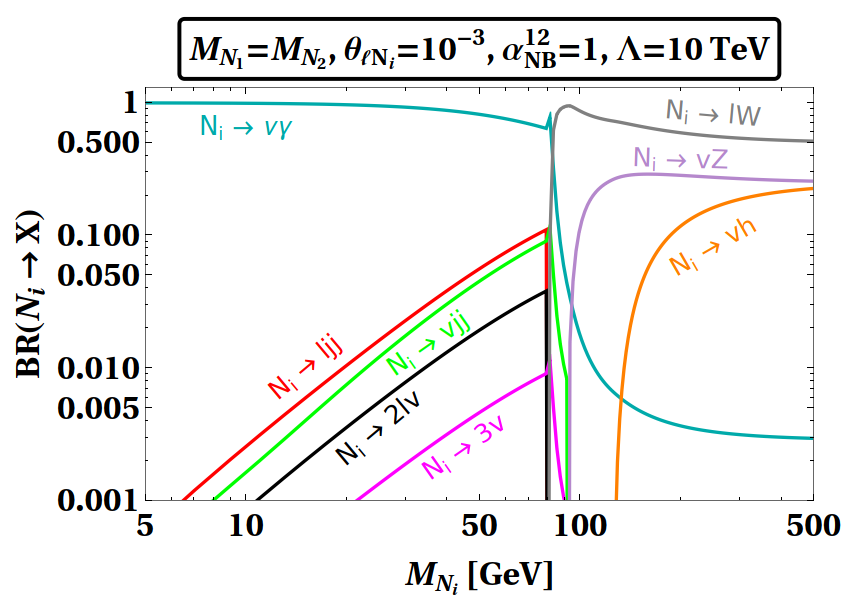}
	\caption{Branching ratios of heavy neutrinos $N_i$ to different final states. The assumption considered for each panel are mentioned in the corresponding boxes. See text for details.}
	\label{fig:BR}
\end{figure}
\par From now on we will assume $\alpha_{NH}=0$ and in this scenario depending on the mass splitting $\delta M_N$, $N_2$ can decay to $N_1\gamma$ and $N_1 Z$ via the $\mathcal{O}_{NB}^5$ operator,
\begin{align}
& \Gamma(N_2\to N_1\gamma) = \frac{2 c_W^2 |\alpha_{NB}^{12}|^2}{\pi \Lambda^2 M_{N_2}^3} (M_{N_2}^2 - M_{N_1}^2 )^3,\\
& \Gamma(N_2\to N_1 Z) = -\frac{s_W^2 |\alpha_{NB}^{12}|^2}{\pi \Lambda^2 M_{N_2}} f_{Z}\left(M_Z, M_{N_2}, M_{N_1}\right)\, \lambda^{\frac{1}{2}}\left(1,\frac{M_{N_1}^2}{M_{N_2}^2}, \frac{M_Z^2}{M_{N_2}^2}\right),
\end{align}
where the functions $f_Z$ and $\lambda$ are defined in Appendix.~\ref{app:decay-widths}. The three-body decay into an off-shell $Z$ boson provides a subdominant contribution to the total decay width. It is clear from the above two equations, the mass splitting $\delta M_N$ and the scale $\Lambda$ are crucial in determining the $N_2$ decay width. In the first three panel of Fig.~\ref{fig:BR}, we show the branching ratios of the heaviest neutrino $N_2$ to various possible final states for different choices of mass splitting $\delta M_N$ with the assumption of $\alpha^{12}_{NB}=1$ and $\Lambda=10$~TeV. We see from the upper left panel of Fig.~\ref{fig:BR} that for relatively small mass splitting $\delta M_N~(\equiv M_{N_2}-M_{N_1})=10$~GeV, the dominant decay mode is $N_2\to N_1\gamma$ up to mass $M_{N_2}\lesssim 200$~GeV, whereas for mass range $M_{N_2}\geq 200$~GeV, two body decay modes~($\ell W$, $\nu Z/h$) coming from the light-heavy mixing $\theta_{\ell N_2}$ starts to dominate. In the case of moderate mass splitting $\delta M_N=50$~GeV, $N_2\to N_1\gamma$ always dominates. In the case of relatively large mass splitting such as $\delta M_N=200$~GeV, in addition to $N_2\to N_1\gamma$ decay mode $N_2\to N_1 Z$ also contributes~\footnote{Note that if the off-diagonal element of $\alpha_{NH}$ is non-zero, then the decay mode $N_2\to N_1 h$ will also contribute along with $N_2\to N_1\gamma$ and $N_2\to N_1 Z$. Although we find that, even with $\alpha_{NB}^{12}\approx \alpha_{NH}^{12}$, $N_2\to N_1\gamma$ will always dominate over others.}.
\begin{figure}[!htbp]
	\centering
	\includegraphics[width=0.45\linewidth]{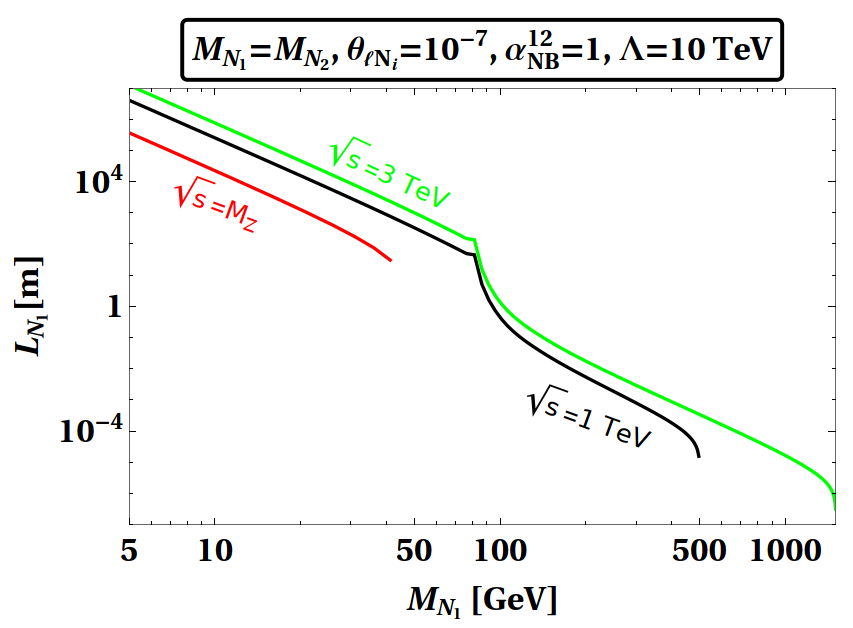}
	\includegraphics[width=0.45\linewidth]{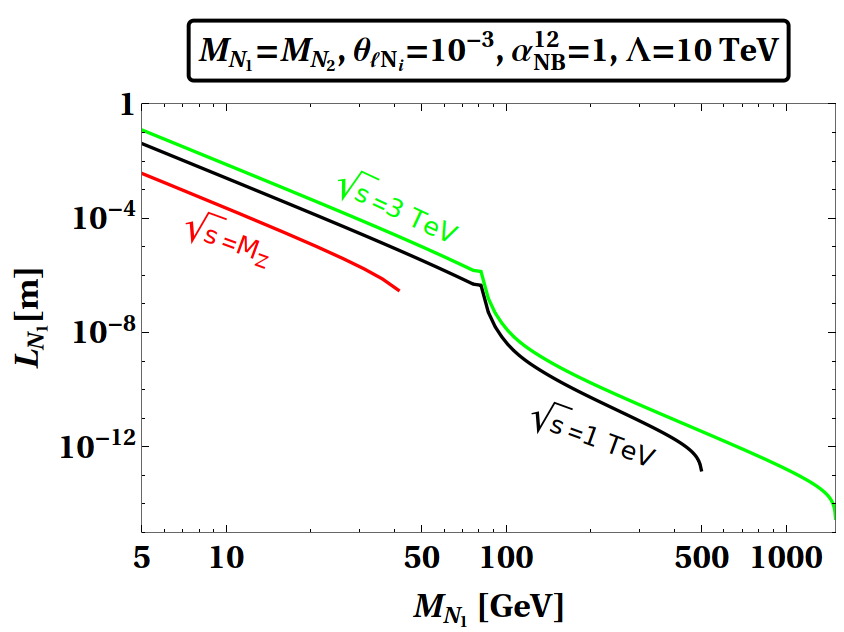}
	\includegraphics[width=0.45\linewidth]{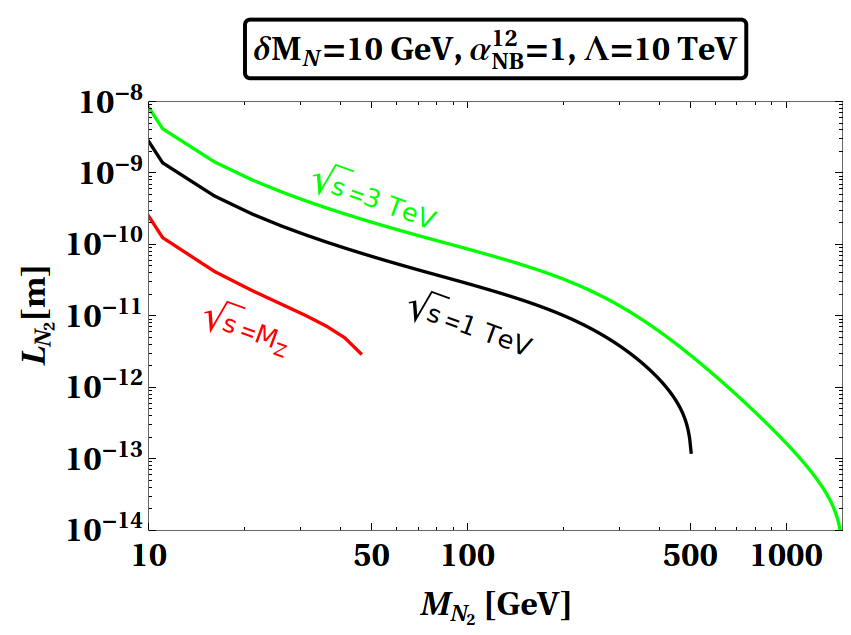}
		\includegraphics[width=0.45\linewidth]{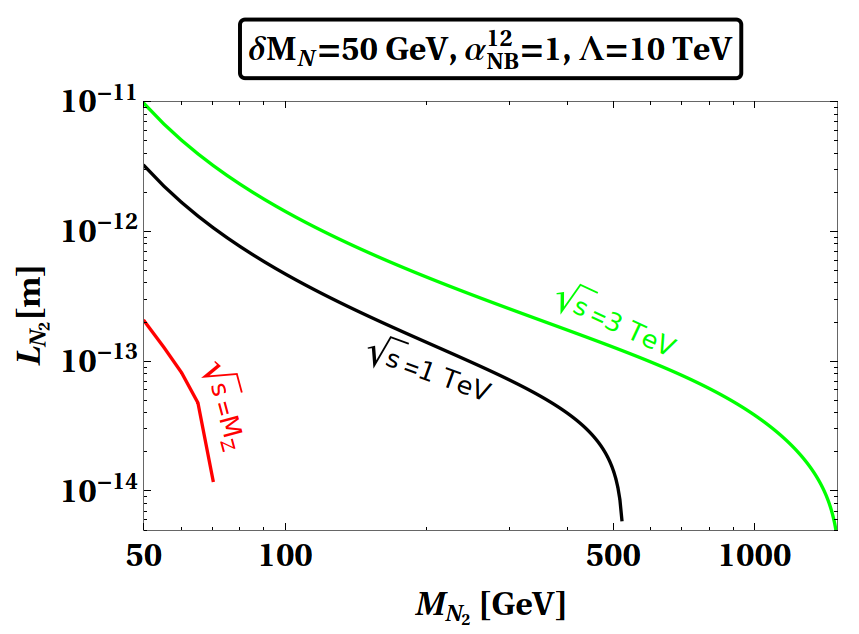}
	\caption{$N_1$ and $N_2$ decay lengths as a function of mass $M_{N_i}$ assuming $N_1$ and $N_2$ is produced through the process $e^+e^-\to N_1 N_2$ at center-of-mass energies $\sqrt{s}=M_Z$~(red), 1 TeV~(black) and 3 TeV~(green).}
	\label{fig:LN}
\end{figure}
\par The lightest RHN $N_1$ on the other hand only decays to SM particles via light-heavy neutrino mixing $\theta_{\ell N_1}$. This implies that $N_1$ decays will always be suppressed by the mixing $\theta_{\ell N_1}$. As the $N_1$ decay widths are proportional to $\theta_{\ell N_1}$, the branching ratios depends weakly on the $\theta_{\ell N_1}$. In the lower right panel of Fig.~\ref{fig:BR}, we show the branching ratios of $N_i$ with the assumptions $M_{N_2}=M_{N_1}$, $\theta_{\ell N_i}=10^{-3}$, $\alpha_{NB}=1$ and $\Lambda=10$~TeV~\footnote{If $N_2$ and $N_1$ are nearly degenerate $M_{N_2}\approx M_{N_1}$, the decays such as $N_2\to N_1\gamma/N_1 Z/N_1 h$ are suppressed and as a result the $N_2$ decay width as well as the branching ratios will be similar to $N_1$.}. In absence of contribution from dimension five operators, heavy neutrinos decay dominantly via three body pure leptonic~($2\ell\nu,3\nu$) or semileptonic mode~($\ell jj, \nu jj$) when $M_{N_i}<M_W$~\cite{Das:2023tna}. But we see that for strong enough magnetic moment interaction, instead of previously mentioned three body decays, two body decay $N_i\to\nu\gamma$ is the dominant mode in the mass range $M_{N_i}<M_W$. Although for relatively large $M_{N_i}$, two body decay modes such as $\ell W$, $\nu Z$ and $\nu h$ coming from pure light-heavy mixing starts to dominate.
\par In Fig.~\ref{fig:LN}, we show the $N_1$ and $N_2$ decay lengths as a function of mass $M_{N_i}$ assuming $N_1$ and $N_2$ are produced through the MMO process $e^+e^-\to N_1 N_2$. Decay lengths are shown for three different center-of-mass energies $\sqrt{s}=M_Z$~(red line), 1 TeV~(black line) and 3 TeV~(green line). We see that the decay lengths of $N_1$ will be large for small mixing or for small mass. If the mixing is $\theta_{\ell N_1}=10^{-3}$, $N_1$ decay lengths will be always small enough to consider the prompt decay, whereas for mixing $\theta_{\ell N_1}=10^{-7}$, $N_1$ is prompt only for masses above 100 GeV. Note that if $M_{N_2}\approx M_{N_1}$, $N_2$ decay length will be same as $N_1$ as long as $\theta_{\ell N_1}\approx \theta_{\ell N_2}$.
\par On the other hand, if the magnetic moment interaction is strong enough and the mass splitting $\delta M_N$ is large enough, the $N_2$ decay length is very small as can be seen in the lower panel of Fig.~\ref{fig:LN}. Hence as long as $M_{N_2}>M_{N_1}$, and $\alpha_{NB}^{12}/\Lambda$ is large enough, $N_2$ decay will always be prompt. Note that in this scenario, dependence on the light-heavy neutrino mixing is mild as the $N_2$ decay is mainly determined by $\alpha_{NB}^{12}$.
\small
\begin{table}[]
\centering
\begin{tabular}{|c||c||c|}
\hline
 Decay chains &  Final state  &  Approximate region of validity \\
\hline
\hline
$ N_{1} \to \nu\gamma$, $N_2\to\nu\gamma$ &  $2\gamma+$Inv &  $10\leq M_{N_i}\leq 70$~GeV, $|\theta|\gsim 10^{-4}$\\
\hline
$N_1\to\ell^\pm_a W^\mp, N_2\to\ell^\pm_b W^\mp, W\to jj $ & $\ell^\pm_a\ell^\pm_b + 4j$ &  $M_{N_i}\geq 100$~GeV, $|\theta|\gsim \theta_{\rm seesaw}$\\
\hline
$ N_{1} \to $ DOD, $N_2\to$ DOD & Inv &  $10\leq M_{N_i}\leq 70$~GeV,  $ |\theta|\lesssim 10^{-4}$\\
\hline
\hline
\end{tabular}
\caption{Different possible collider signatures depending upon the light-heavy neutrino mixing~($\theta_{\ell N_i}$) and heavy neutrino mass~($M_{N_i}$) for the case of degenerate heavy neutrino mass $M_{N_2}=M_{N_1}$. Here $\theta_{\rm seesaw} \sim \sqrt{\frac{m_\nu}{M_{N_i}}}$ and $ N_{i} \to $ DOD represents that $N_i$ decays outside the detector.}
\label{tab:final-states1}
\end{table}
\normalsize
\section{Collider signatures}
\label{sec:collider-signatures}
Having discussed the production and decay modes of the heavy neutrinos, we discuss now the possible collider signatures coming from their production in lepton or hadron colliders: $XX\to N_1 N_2$ where $XX$ stands for either $e^+e^-$ or $pp$. Depending on the model parameters such as the light-heavy mixing, the heavy neutrino mass $M_{N_i}$ and the mass-splitting $\delta M_N$, there will be many interesting collider signatures which we listed in Tables.~\ref{tab:final-states1} and \ref{tab:final-states2}. In the following we discuss them one by one:
\begin{table}[]
\centering
\begin{tabular}{|c||c||c|}
\hline
 Decay chains &  Final state  &  Approximate region of validity \\
\hline
\hline
$N_2\to N_1\gamma$, $N_1\to$ DOD & $\gamma+$Inv & $10\leq M_{N_1}\leq 70$~GeV, $|\theta|\lesssim 10^{-4}$ \\
\hline
$N_2\to N_1\gamma$, $N_1\to\nu\gamma$ & $3\gamma+$Inv &   $10\leq M_{N_1}\leq 70$~GeV, $|\theta|\gsim 10^{-4}$ \\
\hline
$N_2\to N_1\gamma$, $N_1\to\ell^\pm_a W^\mp$, $W^\mp\to jj$ & $\ell^\pm_a\ell^\pm_a + 4j + \gamma $ &  $M_{N_1}\geq 100$~GeV, $|\theta|\gsim \theta_{\rm seesaw}$ \\
\hline
\hline
\end{tabular}
\caption{Different possible collider signatures depending upon the light-heavy neutrino mixing~($\theta_{\ell N_i}$), heavy neutrino mass~($M_{N_i}$) and mass splitting $\delta M_{N}\geq 10$~GeV.  Here, $\theta_{\rm seesaw} \sim \sqrt{\frac{m_\nu}{M_{N_i}}}$ and $ N_{1} \to $ DOD represents that $N_1$ decays outside the detector.}
\label{tab:final-states2}
\end{table}
\begin{itemize}
\item {\bf $2\gamma+\text{Inv}$: }
If the heavy neutrino masses are degenerate and lies in the range $10 \text{ GeV }\leq M_{N_i} \leq 70\text{ GeV}$, then from Fig.~\ref{fig:BR} we see that both heavy neutrinos $N_1$ and $N_2$ dominantly decays to $\nu\gamma$. This decay chain can give us the final state $2\gamma+\text{Inv}$, e.g,
\begin{align}
XX\to N_1 N_2, \,\, N_{1,2}\to\nu\gamma \Longrightarrow 2\gamma+\text{Inv}.
\end{align}
This signal is of course not valid for all values of mixing $\theta$. Depending on the value of $\theta$, $N_i$ decays inside or outside the detector. The probability of the heavy neutrinos $N_i$ to decay  after travelling a distance $\ell_D$ from the primary vertex is given by,
\begin{align}
\mathcal{P}^{N_i}_{\rm dec}(\ell_D)=1-\text{exp}\left(-\ell_D \Gamma_{N_i}\frac{M_{N_i}}{p_{N_i}}\right)=1-\text{exp}\left(-\frac{\ell_D}{\ell_{N_i}}\right),
\end{align}
where $\ell_{N_i}=p_{N_i}/(M_{N_i}\Gamma_{N_i})$ is the $N_i$ decay length with $p_{N_i}$ is the three momentum of $N_i$ coming from specific production mechanism. Then the production rates for this signal can be written as
\begin{align}
\sigma^{\rm NP}(XX\to 2\gamma + \text{Inv})=\sigma(XX\to N_1 N_2) \text{BR}(N_2\to \nu\gamma) \text{BR}(N_1\to \nu\gamma) \epsilon_{\rm cuts} \epsilon_{\rm det} \mathcal{P}^{N_2}_{\rm dec}(\ell_D) \mathcal{P}^{N_1}_{\rm dec}(\ell_D),
\end{align}
where $\epsilon_{\rm cuts}$ and $\epsilon_{\rm det}$ are the efficiencies of the kinematic cuts and detection for the final photon. Note that for degenerate heavy neutrino masses, we have $\mathcal{P}^{N_2}_{\rm dec}(\ell_D)=\mathcal{P}^{N_1}_{\rm dec}(\ell_D)$. We find  $\mathcal{P}^{N_i}_{\rm dec}(\ell_D)\approx 1$ for the mass range $10 \text{ GeV }\leq M_{N_i} \leq 70\text{ GeV}$ only if $|\theta|>10^{-4}$.
\begin{figure}[!htbp]
	\centering
	\includegraphics[width=0.45\linewidth]{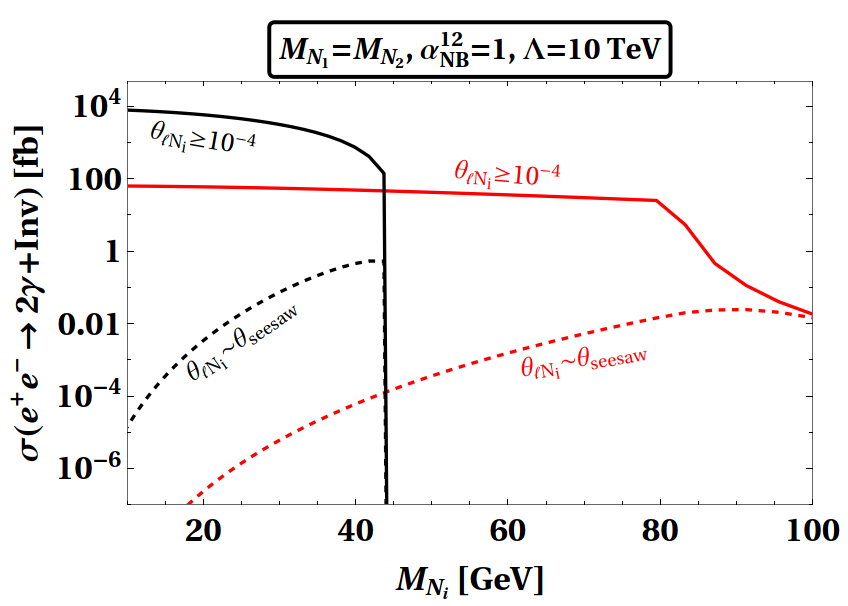}
	\includegraphics[width=0.45\linewidth]{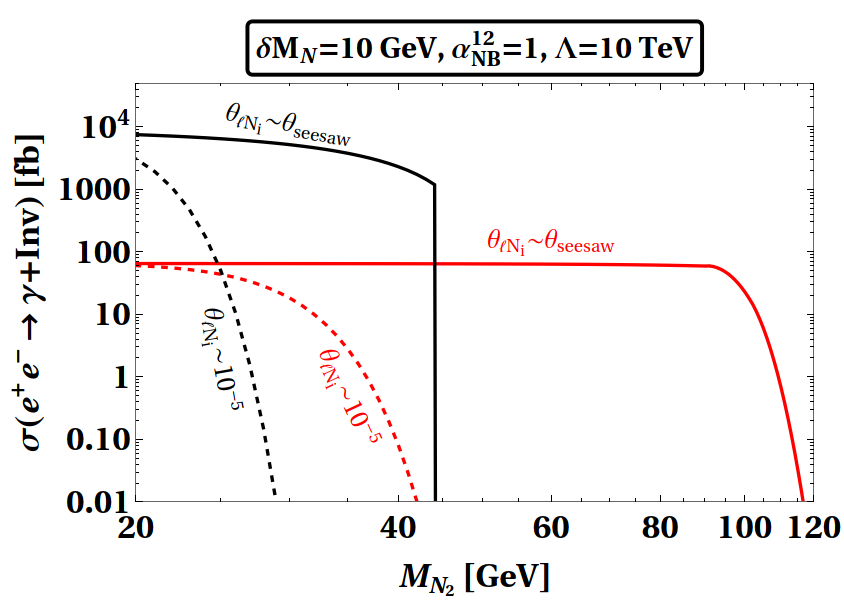}
	\includegraphics[width=0.45\linewidth]{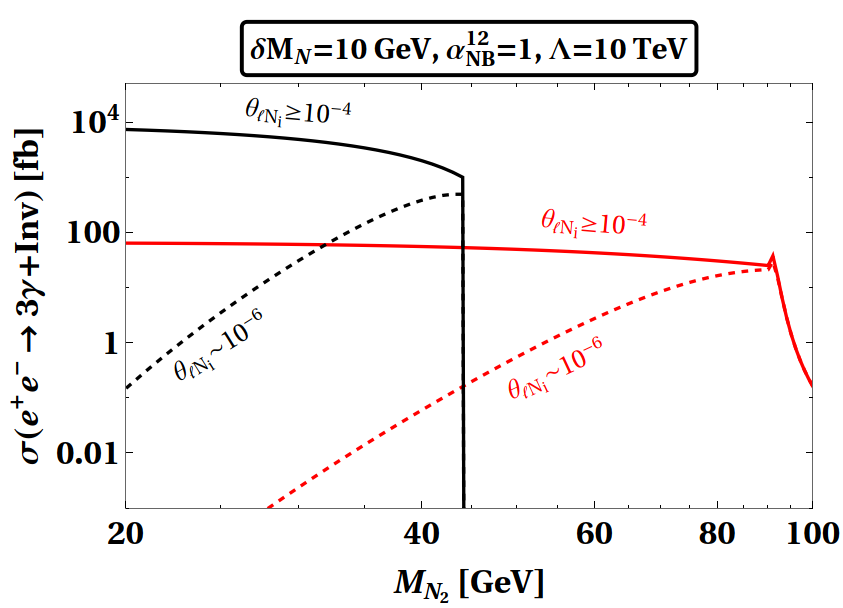}
	\caption{Production cross section at lepton collider for the processes such as $2\gamma+\text{Inv}$~(upper left panel), $\gamma+\text{Inv}$~(upper right panel) and $3\gamma+\text{Inv}$~(lower panel). The black and red lines stand for two center-of-mass energies, $\sqrt{s}=M_Z$ (black) and $\sqrt{s}=1$~TeV (red), respectively. The solid and dashed lines stand for different values of mixing. See main text for details.}
	\label{fig:gammaSignal}
\end{figure}
In the upper left panel of Fig.~\ref{fig:gammaSignal} we show the production cross-section for this signal at lepton collider for two benchmark values of mixing $|\theta_{\ell N_i}|>10^{-4}$~(solid lines) and $|\theta_{\ell N_i}|\sim \theta_{\rm seesaw}$~(dashed lines), respectively~\footnote{One important point to note is that when we are showing the cross-section, we have considered the idealized detector with $100\%$ detection, reconstruction efficiencies etc.}. The black and red lines stand for two center-of-mass energies, $\sqrt{s}=M_Z$ and $\sqrt{s}=1$ TeV, respectively. Comparing the dashed with solid lines we see that the production cross-section drops sharply for small mixing. This is due to the fact that for small mixing and degenerate masses, the probability for both the heavy neutrinos to decay inside the detector is very small, $\mathcal{P}^{N_{1,2}}_{\rm dec}(\ell_D)\ll 1$.

\item {\bf $\gamma+\text{Inv}$: }
The other possible collider signatures is the event with single photon in association with missing energy. This happens when magnetic moment is strong enough but the light-heavy neutrino mixing is small such that $N_2\to N_1\gamma$ is prompt but $N_1$ decays outside the detector~(DOD), e.g,
\begin{align}
XX\to N_1 N_2, \,\, N_2\to N_1\gamma, \,\, N_1 \to \text{DOD} \Longrightarrow \gamma+\text{Inv}.
\label{eq:ttt}
\end{align}
This signal is valid in the region $\delta M_N > 10$~GeV and $M_{N_1}<70$~GeV. The relative mass splitting~($\delta M_N$) is very important for validity of this final state. In the $N_2$ rest frame, photon energy is estimated as
\begin{align}
E_\gamma=\frac{M_{N_2}+M_{N_1}}{2 M_{N_2}}\delta M_N.
\end{align}
We see that smaller the mass splitting $\delta M_N$ the softer the final state photon coming from $N_2\to N_1\gamma$ decay, which, in order to be identified by the corresponding detector, must satisfy a minimal threshold criterion. We find that for our interested mass range of $M_{N_1}$, choosing $\delta M_N>10$ GeV one can always make sure the minimum threshold criterion of $E_\gamma>10$ GeV which is the photon acceptance cut for most of the detector. Keeping this in mind we can write the production rates for the above mentioned signal~(Eq.~\ref{eq:ttt}) can be written as
\begin{align}
\sigma^{\rm NP}(XX\to \gamma + \text{Inv})=\sigma(XX\to N_1 N_2) \text{BR}(N_2\to N_1\gamma) \epsilon_{\rm cuts} \epsilon_{\rm det} \mathcal{P}^{N_2}_{\rm dec}(\ell_D) \Big(1-\mathcal{P}^{N_1}_{\rm dec}(\ell_D)\Big)^2,
\end{align}
where $\Big(1-\mathcal{P}^{N_1}_{\rm dec}(\ell_D)\Big)$ stands for the $N_1$'s probability to decay outside the detector. Hence this factor should be large to observe this final state. Note that we have $\text{BR}(N_2\to N_1\gamma)\approx 1$ as long as $\delta M_N>10$~GeV for our interested range of parameters. In the upper right panel of Fig.~\ref{fig:gammaSignal}, we show the production cross-section for this signal again for two values of the mixing, $|\theta_{\ell N_i}|\sim \theta_{\rm seesaw}$~(solid lines) and $|\theta_{\ell N_i}|\sim 10^{-5}$~(dashed lines), respectively. The black and red lines stand for two center-of-mass energies, $\sqrt{s}=M_Z$ and $\sqrt{s}=1$ TeV, respectively. Comparing the dashed with solid lines we see that the production cross-section drops sharply when mixing is relatively large, $|\theta_{\ell N_i}|>10^{-5}$. This is due to the fact that for comparatively large mixing, the probability for the heavy neutrino $N_1$ to decay inside the detector is large, i.e,  $\Big(1-\mathcal{P}^{N_1}_{\rm dec}(\ell_D)\Big)\ll 1$.
\item {\bf $3\gamma+\text{Inv}$: } If the heavy neutrino masses are non-degenerate~($\delta M_N>10$~GeV) and lies in the range $10 \text{ GeV }\leq M_{N_i} \leq 70\text{ GeV}$, then we can have following dominant decay chains $N_2\to N_1\gamma$ and $N_1\to\nu\gamma$. With this decay chain we have the following final state $3\gamma + \text{Inv}$, e.g,
\begin{align}
XX\to N_1 N_2, \,\, N_2\to N_1\gamma, \,\, N_1 \to \nu\gamma \Longrightarrow 3\gamma+\text{Inv}.
\end{align}
Note that this signal is approximately valid if $|\theta_{\ell N_i}|>10^{-4}$ as otherwise $N_1$ will almost always decay outside the detector. The production rates for this signal can be written as
\small
\begin{align}
\sigma^{\rm NP}(XX\to 3\gamma + \text{Inv})=\sigma(XX\to N_1 N_2) \text{BR}(N_2\to N_1\gamma) \Big(\text{BR}(N_1\to \nu\gamma)\Big)^2 \epsilon_{\rm cuts} \epsilon_{\rm det} \mathcal{P}^{N_2}_{\rm dec}(\ell_D) \Big(\mathcal{P}^{N_1}_{\rm dec}(\ell_D)\Big)^2,
\end{align}
\normalsize
\begin{figure}[!htbp]
	\centering
	\includegraphics[width=0.45\linewidth]{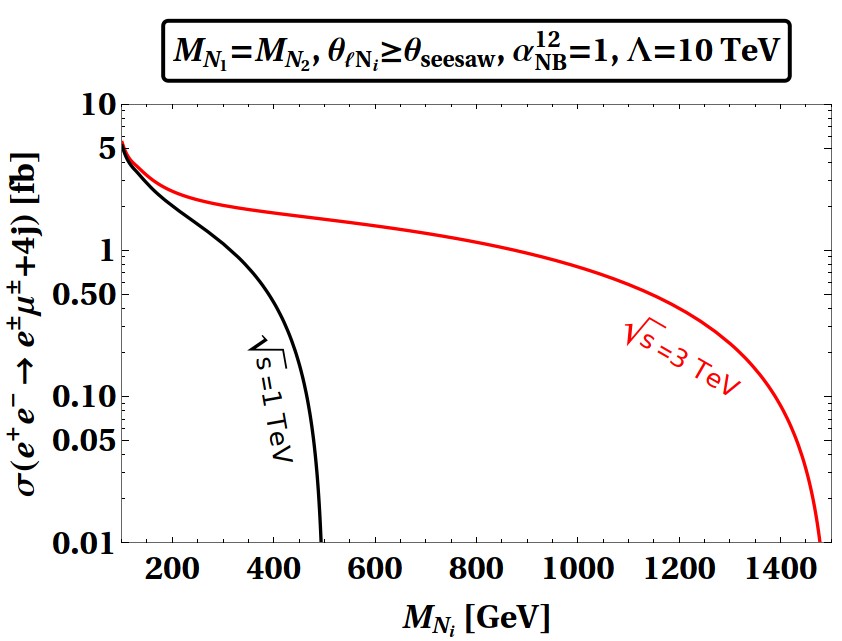}
	\includegraphics[width=0.45\linewidth]{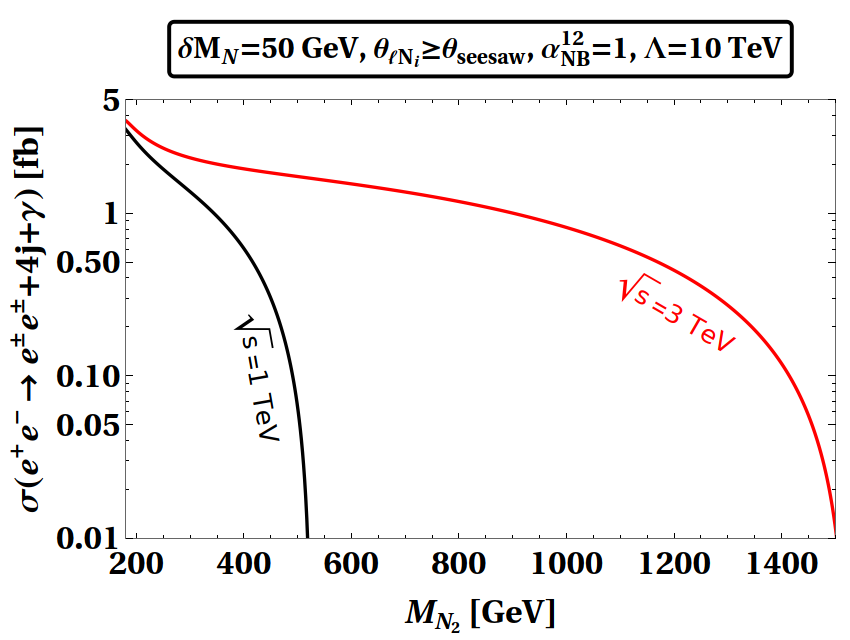}
	\caption{Production cross section at lepton collider for the lepton number violating processes such as $e^\pm \mu^\pm 4j$~(left panel) and $e^\pm e^\pm 4j+\gamma$~(right panel). The black and red lines stand for two center-of-mass energies, $\sqrt{s}=1$ TeV (black) and $\sqrt{s}=3$~TeV (red), respectively. See the main text for details.}
	\label{fig:leptonSignal}
\end{figure}
In the lower panel of Fig.~\ref{fig:gammaSignal}, we show the production cross-section for this signal again for two values of mixing, $|\theta_{\ell N_i}| \geq 10^{-4}$~(solid lines) and $|\theta_{\ell N_i}|\sim 10^{-6}$~(dashed lines), respectively. Similar to the above cases, the black and red lines stand for two center-of-mass energies, $\sqrt{s}=M_Z$ and $\sqrt{s}=1$ TeV, respectively. We see that the production cross-section drops sharply when mixing is relatively small~($|\theta_{\ell N_i}|\leq 10^{-5}$) as the probability of $N_1$ decaying inside the detector becomes small, $\mathcal{P}^{N_1}_{\rm dec}(\ell_D) \ll 1$.
\item {\bf $\ell_a^\pm\ell_b^\pm + 4j$: }
If the heavy neutrino masses are degenerate and $M_{N_i}>100$~GeV then both the heavy neutrinos dominantly decays to $\ell W$, $\nu Z$ or $\nu h$ through the mixing $\theta$~(see Fig.~\ref{fig:BR}). In this case considering the decay chain $N_i\to\ell_a^\pm W^\mp,\, W^\mp \to j j$ the $N_1 N_2$ production gives the following final states,
\begin{align}
XX\to N_1 N_2, \,\, N_1\to \ell_a^\pm j j, \,\, N_2 \to \ell_b^\pm j j \Longrightarrow \ell_a^\pm\ell_b^\pm + 4j.
\end{align}
Similarly, the production rates for this signal can be written as
\begin{align}
\sigma^{\rm NP}(XX\to \ell_a^\pm\ell_b^\pm + 4j)=\sigma(XX\to N_1 N_2) \text{BR}(N_1\to \ell_a^\pm j j) \text{BR}(N_2\to \ell_b^\pm j j) \epsilon_{\rm cuts} \epsilon_{\rm det} \mathcal{P}^{N_2}_{\rm dec}(\ell_D) \mathcal{P}^{N_1}_{\rm dec}(\ell_D),
\end{align}
Note that as long as $|\theta|>\theta_{\rm seesaw}$ and $M_{N_i}>100$~GeV we find $\mathcal{P}^{N_i}_{\rm dec}(\ell_D)\approx 1$. This final states clearly violates lepton number by two units and hence the expected SM background is small. Note that if both of the like sign dilepton~($\ell_a^\pm \ell_b^\pm$) are not of the same flavor~($\ell_a\neq\ell_b$), then the process is not only lepton number violating but also lepton flavor violating. In the left panel of Fig.~\ref{fig:leptonSignal}, we show the production cross-section for this signal for two values of center-of-mass energy, $\sqrt{s}=1$ TeV~(black line) and $\sqrt{s}=3$ TeV~(red line), respectively.
\item {\bf $\ell_a^\pm\ell_b^\pm + 4j+\gamma$: }
If the heavy neutrino masses are non-degenerate $\delta M_N>10$~GeV and magnetic moment is strong enough then $N_2\to N_1\gamma$ decay dominates. Furthermore, if $M_{N_1}>100$~GeV, one can consider the decay chain $N_1\to\ell^{\pm}jj$. This can give us the following final state,
\begin{align}
XX\to N_1 N_2, \,\, N_2\to N_1\gamma, \,\, N_1\to \ell_a^\pm j j,\,\, N_1\to \ell_b^\pm j j \Longrightarrow \ell_a^\pm\ell_b^\pm + 4j+\gamma
\end{align}
The production rates for this signal can be written as
\small
\begin{align}
\sigma^{\rm NP}(XX\to \ell_a^\pm\ell_b^\pm + 4j+\gamma) & =\sigma(XX\to N_1 N_2) \text{BR}(N_2\to N_1\gamma) \text{BR}(N_1\to \ell_a^\pm j j) \text{BR}(N_1\to \ell_b^\pm j j)\nonumber \\
& \epsilon_{\rm cuts} \epsilon_{\rm det} \mathcal{P}^{N_2}_{\rm dec}(\ell_D) \Big(\mathcal{P}^{N_1}_{\rm dec}(\ell_D)\Big)^2,
\end{align}
\normalsize
If $\delta M_N$ and magnetic moment is relatively large then $\mathcal{P}^{N_2}_{\rm dec}(\ell_D)\approx 1$. Also as long as $|\theta|>\theta_{\rm seesaw}$ and $M_{N_1}>100$~GeV we find $\mathcal{P}^{N_1}_{\rm dec}(\ell_D)\approx 1$. This final state also violates lepton number by two units and hence the SM background is expected to be very much suppressed. In the right panel of Fig.~\ref{fig:leptonSignal}, we show the production cross-section for this signal for two values of center-of-mass energy, $\sqrt{s}=1$ TeV~(black line) and $\sqrt{s}=3$ TeV~(red line), respectively.
\end{itemize}
\par Looking for the displaced decays may be another important method of searching for RHNs. For example, rather than looking for events with prompt leptons, one can look for events with displaced leptons coming from relatively long-lived heavy neutrinos, e.g,
\begin{align}
& XX\to N_1 N_2, \,\,\,\,\, \text{Displaced: } N_2/N_1\to \ell_a^\pm\ell_b^\mp\nu/\ell_a^\pm j j,\\
& XX\to N_1 N_2, \, \, \text{Prompt: } N_2\to N_1\gamma, \,\, \text{Displaced: } N_1\to \ell_a^\pm\ell_b^\mp\nu/\ell_a^\pm j j.
\end{align}
The problem is that for $M_{N_i}>100$~GeV, heavy neutrinos will be mostly short-lived even for very small mixing $\theta_{\ell N_i}\sim \theta_{\rm seesaw}$. On the other hand for $M_{N_i}<100$~GeV, heavy neutrinos can be long-lived for relatively small mixing but for $M_{N_i}<100$~GeV, the $\text{BR}(N_i\to 2\ell\nu)$ or $\text{BR}(N_i\to \ell 2j)$ itself very small~(see lower right panel of Fig.~\ref{fig:BR}). Due to this reason we have not discussed these type of displaced decay further~\footnote{Note that without dimension five operator, $N_i\to 2\ell\nu/\ell 2j$ decays are dominated by mixing and in that case branching ratios $\text{BR}(N_i\to 2\ell\nu/\ell 2j)$ can be large for $M_{N_i}<100$~GeV. This displaced vertex scheme has been used in Refs.~\cite{CMS:2022fut,ATLAS:2022atq} at CMS and ATLAS to look for RHNs with mixing only.}.
\section{Current collider limits and future sensitivity}
\label{sec:constraints}
\subsection{Existing Collider constraints}
In this section we discuss several existing constraints on the MMO. The model under consideration in this paper has three parameters: the mass splitting $\delta M_N$, the heavy neutrino mass $M_{N_1}$, and the Wilson coefficient of the MMO, which is parametrized by $\Lambda$ assuming $|\alpha_{NB}^{12}|=1$. This parameter space is already constrained by various existing laboratory data from colliders and we discuss here most relevant of them. There are analyses to search for monophoton/diphoton in association of missing energy  at LEP and LHC. For $\gamma+\text{Inv}$, we consider LEP1 data at the center-of-mass energy around the $Z$, and LEP2 data at larger center-of-mass energy. For $2\gamma+\text{Inv}$, LEP2 data will be used. There are also LHC searches for both of these signals. Additional $Z$ boson decay modes can arise from the MMO such as $Z\to\gamma+\text{Inv}$, $2\gamma+\text{Inv}$ and $Z\to\text{Inv}$. The $Z$ boson decay width has been measured accurately by LEP and hence can constrain the MMO. 
\begin{figure}[b]
	\centering
 \includegraphics[width=0.45\linewidth]{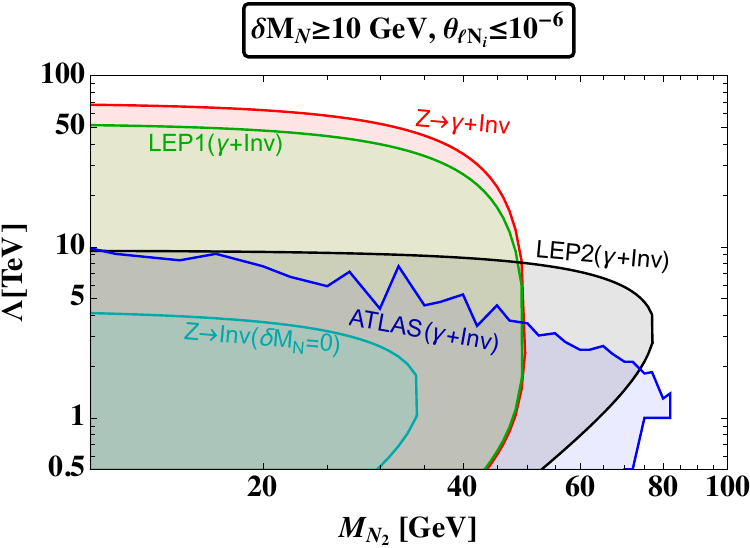}
 \includegraphics[width=0.45\linewidth]{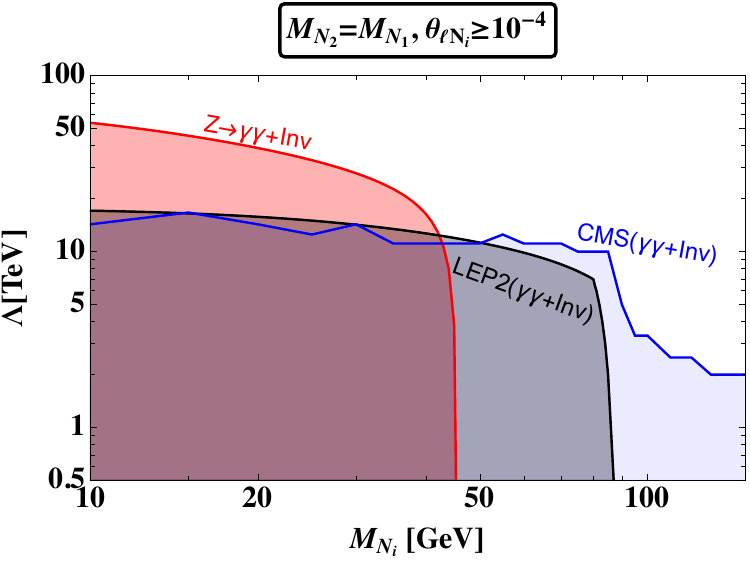}
	\caption{Excluded region in the $M_N-\Lambda$ plane from LEP and LHC searches. The left panel is for mostly non-degenerate heavy neutrino masses~(apart from $Z\to\text{Inv}$) and relatively small mixing~($|\theta_{\ell N_i}|\leq 10^{-6}$), whereas the right panel is for degenerate heavy neutrino masses and relatively large mixing~($|\theta_{\ell N_i}|>10^{-4}$).}
	\label{fig:BoundLEP}
\end{figure}
\begin{itemize}
\item  \underline{Z decays from LEP:} The presence of MMO will affect the $Z$ boson decays which are examined precisely by the LEP experiments~\cite{OPAL:1993ezs}. In the case of degenerate heavy neutrinos~($M_{N_2}=M_{N_1}$) and $M_{N_i}< M_Z/2$, $Z$ can decay to $2\gamma+\text{Inv}$ and its rate is given by
\begin{align}
\Gamma_{Z\to 2\gamma + \text{Inv}}=\Gamma_{Z\to N_1 N_2} \text{BR}(N_1\to\nu\gamma) \text{BR}(N_2\to\nu\gamma)\epsilon_{\rm cuts} \mathcal{P}^{N_1}_{\rm dec}(\ell_D)  \mathcal{P}^{N_2}_{\rm dec}(\ell_D),
\end{align}
where $\epsilon_{\rm cuts}$ is the efficiency of the kinematic cuts for the photons and the expression for the $Z\to N_1 N_2$ decay width is
\begin{align}
\Gamma_{Z\to N_1 N_2}=\frac{2s_W^2}{3\pi \Lambda^2 M_Z}|\alpha_{NB}^{12}|^2 f_{Z}\left(M_Z, M_{N_1}, M_{N_2}\right)\lambda^{\frac{1}{2}}\Big(1,\frac{M_{N_1}^2}{M_Z^2},\frac{M_{N_2}^2}{M_Z^2}\Big),
\end{align}
where the expressions for functions $\lambda$ and $F_Z$ are given in Appendix.~\ref{app:decay-widths}. Absence of this signal $2\gamma+\text{Inv}$ at LEP1~\cite{OPAL:1993ezs} set an upper limit on the branching ratio $\text{BR}(Z\to 2\gamma + \text{Inv})<3.1\times 10^{-6}$ at $95\%$ C.L. On the other hand if the heavy neutrinos are non-degenerate~($M_{N_2}>M_{N_1}$) and $M_{N_1}+M_{N_2}<M_Z$, then the signal of $\gamma+\text{Inv}$ can arise from the $Z$ decay whose rate is given by 
\begin{align}
\Gamma_{Z\to \gamma + \text{Inv}}=\Gamma_{Z\to N_1 N_2} \text{BR}(N_2\to N_1\gamma) \epsilon_{\rm cuts} \mathcal{P}^{N_2}_{\rm dec}(\ell_D)  (1-\mathcal{P}^{N_1}_{\rm dec}(\ell_D))^2.
\end{align}
LEP1~\cite{L3:1997exg} set an upper limit on the branching ratio for this final state: $\text{BR}(Z\to \gamma + \text{Inv})<3.2\times 10^{-6}$ at $95\%$ C.L. Besides the above two final states, the invisible decay of the $Z$ boson can occur if heavy neutrinos are degenerate and decay outside the detector,
\begin{align}
\Gamma_{Z\to \text{Inv}}=\Gamma_{Z\to N_1 N_2} \left(1-\mathcal{P}^{N_1}_{\rm dec}(\ell_D)\right) \left(1-\mathcal{P}^{N_2}_{\rm dec}(\ell_D)\right).
\end{align}
LEP experiments place a strong bound on the new physics contributions to the $Z$-boson invisible decay width as $\Gamma^{\rm NP}_{Z\to \text{Inv}}<2$~MeV at $95\%$ C.L~\cite{ALEPH:2005ab}. The exclusion region in $M_N-\Lambda$ plane~(assuming $\alpha_{\rm NB}^{12}=1$) coming from $Z\to\text{Inv}$ and $Z\to\gamma+\text{Inv}$ are shown in the left panel of Fig.~\ref{fig:BoundLEP}, whereas for $Z\to 2\gamma+\text{Inv}$ is shown in the right panel. We show the constraints coming from $Z\to \text{Inv}/\gamma+\text{Inv}$ and $Z\to 2\gamma+\text{Inv}$ in two different panel as their region of validity are different. For example, the constraints coming from $Z\to\text{Inv}$~($\gamma+\text{Inv}$) are only relevant when $|\theta_{\ell N_i}|\leq 10^{-6}$ as then only $N_{1,2}$~($N_1$) will decay outside the detector giving the missing energy signature, whereas for $Z\to 2\gamma+\text{Inv}$ both $N_1$ and $N_2$ needs to decay inside the detector, hence only valid for large mixing $|\theta_{\ell N_i}|\geq 10^{-4}$. Note that some parts of the exclusion region may not be meaningful in the sense that they lies outside of the EFT validity region. For the EFT to be valid we require $M_{N_i},\,\, \sqrt{s} < \Lambda$, where $s$ is the Lorentz invariant energy that enters the vertex for $N_1 N_2$ production.
\item  \underline{$\gamma + \text{Inv}$ search at LEP:}
There have been many analyses dedicated for the $\gamma+\text{Inv}$ final state at LEP~\cite{L3:1992cmn,DELPHI:1996drf,DELPHI:2003dlq,OPAL:1994kgw}. We decide to use the results of LEP1~($\sqrt{s}=M_Z$) and  LEP2 ($\sqrt{s}=200\sim 209$ GeV). LEP1 sets an upper bound of 2 pb~(0.15 pb) on the cross section of new physics contribution to the $\gamma+\text{Inv}$ final state with the requirements of $|\cos\theta_\gamma|<0.7$ and $E_\gamma>2$~GeV~(23 GeV) (see Fig.~12 of Ref.~\cite{OPAL:1994kgw}). We find that the selection cut efficiency $\epsilon_{\rm cut}$ varies in the range $70\%-80\%$~($40\%-50\%$) for $E_\gamma > 2$ GeV~($E_\gamma > 23$ GeV). For LEP2, we use DELPHI data at $\sqrt{s}=200\sim 209$ GeV in the angular region of $45^{\degree} < \theta_\gamma < 135^{\degree}$ and $0.06<x_\gamma<1.1$ with $x_\gamma=2E_\gamma/\sqrt{s}$~\cite{DELPHI:2003dlq}. This sets an upper bound of 0.05 pb and we find the detection efficiency to be around $\epsilon_{\rm cut}=65\%$. The corresponding exclusion region for LEP1~(green shaded) and LEP2~(gray shaded) are shown in the left panel of Fig.~\ref{fig:BoundLEP}. This constraint is valid only when $|\theta_{\ell N_i}|\leq 10^{-6}$ as then only $N_1$ will decay outside the detector giving the missing energy signature.
\item \underline{$2\gamma + \text{Inv}$ search at LEP:} There is also dedicated analysis for the final state $2\gamma + \text{Inv}$ at LEP2 with $\sqrt{s}\sim 207$~GeV~\cite{L3:2003yon}.
The new physics contribution to the observed  $2\gamma + \text{Inv}$ signal is constrained as $ \sigma^{\rm NP}\le 0.02$~pb (see third panel of Fig.~6 of Ref.~\cite{L3:2003yon}). The selection criteria for this are as follows: two photons with $E_\gamma>1$~GeV, for most energetic photon $14^{\degree}< \theta_{\gamma 1}< 166^{\degree}$ and for the  second one $11^{\degree}< \theta_{\gamma 2}< 169^{\degree}$ and transverse momentum of the two photon system $p_T^{\gamma\gamma}>0.02 \sqrt{s}$. We obtain the selection cut efficiency $\epsilon_{\rm cut}>90\%$  for considered range of $M_{N_{i}}$. The corresponding exclusion region is shown in gray shaded region in the right panel of Fig.~\ref{fig:BoundLEP}. This constraint is valid for $10\text{ GeV}\leq M_{N_i}\leq 70\text{ GeV}$ and $|\theta_{\ell N_i}|\geq 10^{-4}$.
\item \underline{$pp\to 2\gamma+\text{Inv}$ search  at LHC: }
The CMS collaboration has searched for $2\gamma+\text{Inv}$ signal using $35.9~\text{fb}^{-1}$ data at 13 TeV center-of-mass energy~\cite{CMS:2019vzo}. This search can be used to indirectly set constraints on the process  $pp\to N_1 N_2\to 2\gamma+\text{Inv}$. The Ref.~\cite{Biekotter:2020tbd} has already recast this CMS search to set limits on the allowed BSM events in different bins of  missing energy distribution. We calculated the $pp\to N_1 N_2\to 2\gamma+\text{Inv}$ events using the cuts: $N^\gamma \ge2$, $p_T^\gamma > 40$ GeV, $|\eta^\gamma|<1.44$, $\text{MET}>100$ GeV, $M_{\gamma\gamma}>105$ GeV and $\Delta R(\gamma\gamma) >0.6$.  We set a limit on $\Lambda$ comparing the signal events with the $s_{\text{max}}$ given in Table.~\ref{tab:smax} (taken from Table.~4 of Ref.~\cite{Biekotter:2020tbd}) for specific bins of MET distribution. This is shown in the right panel of Fig.~\ref{fig:BoundLEP}~(blue shaded region) and valid only for $|\theta_{\ell N_i}|\geq 10^{-4}$.
\begin{table}[]
	\centering
	\begin{tabular}{|c|c|c|c|c|c|}
		\hline
$MET$		& [100-115] GeV &[115-130] GeV & [130-150] GeV &[150-185] GeV  &[185-250] GeV  \\ \hline
 $s_{\text{max}}$&23.5  &15.0  &10.5  &14.3  &9.6 \\ \hline
	\end{tabular}
\caption{ The maximum number of allowed additional signal events $s_{\text{max}}$ in various $MET$ bins.}
\label{tab:smax}
\end{table}
\item \underline{$\gamma + \text{Inv}$ search at LHC:}
The ATLAS collaboration has searched for dark matter in the final states containing a photon and
missing transverse momentum \cite{ATLAS:2020uiq} which is applicable to the process $pp\to N_1 N_2\to \gamma+\text{Inv}$. This search used $139~\text{fb}^{-1}$ data at 13 TeV center-of-mass energy. We follow the selection cuts provided in Table.~2 of Ref.~\cite{ATLAS:2020uiq} which are $N^\gamma \ge1$, for leading photon $p_T^\gamma > 150$ GeV, $|\eta^\gamma|<1.37$ or $1.52 < |\eta^\gamma|<2.37$, and $\Delta \Phi(\gamma,MET) >0.4$. The observed upper limits at $95\%$ C.L. on the  BSM signal events ($N_{\text{obs}}^{95}$) for various signal regions are reported in Table.~7 of Ref.~\cite{ATLAS:2020uiq}. We consider the three signal regions SRE1, SRE2 and SRE3 with $\text{MET}$ in the ranges $[200-250],~ [250-300],~ [300-375]$ GeV and the respective $N_{\text{obs}}^{95}$, 250, 145 and 109. The resulting constraint is shown in the left panel of Fig.~\ref{fig:BoundLEP}~(blue shaded region) and valid only for $|\theta_{\ell N_i}|\leq 10^{-6}$.
\end{itemize}
The upshot of the above discussion is that for non-degenerate case~($\delta M_N>10$ GeV) the tightest bound on the cut-off scale $\Lambda\gtrsim 70$ TeV in the mass range $10\text{ GeV}\lesssim M_{N_2}\lesssim 40$ GeV comes from the signal $Z\to\gamma+\text{Inv}$ at LEP, whereas for the mass range $40\text{ GeV}\lesssim M_{N_2}\lesssim 70$ GeV, the tightest limit on the cut-off scale $\Lambda\gtrsim 10$ TeV comes from the final state $e^+e^-\to\gamma+\text{Inv}$ at LEP2. In the degenerate case, the tightest bound in the mass range $10\text{ GeV}\lesssim M_{N_i}\lesssim 40$ GeV comes from $Z\to\gamma\gamma+\text{Inv}$ at LEP1 and for higher mass range $M_{N_i}\gtrsim 40$ GeV tightest bound comes from $\gamma\gamma+\text{Inv}$ at LEP2 or CMS.
\par Note that if the heavy neutrino masses are of the order of few GeV, they can be produced not only at high energy colliders, but also at fixed target experiments via meson decay, $M\to N_1 N_2$~\cite{Barducci:2022gdv}. Subsequent heavier neutrino decay $N_2\to N_1\gamma$ give rise to single-$\gamma$ events, which can be detected by experiments with background controlled environment. Ref.~\cite{Barducci:2022gdv} computed the current bound at CHARM~\cite{CHARM:1985anb}, NuCal~\cite{Blumlein:1990ay,Blumlein:1991xh} and NA64~\cite{Bernhard:2020vca} as well as the future sensitivity at ANUBIS~\cite{Bauer:2019vqk}, CODEX-b~\cite{Gligorov:2017nwh,Aielli:2022awh}, FASER2~\cite{Feng:2017uoz,Feng:2022inv} and SHiP~\cite{SHiP:2015vad,SHiP:2021nfo}. They found that tightest limit on the cut-off scale~($\Lambda\sim 10^5$ GeV) comes from SHiP.
\subsection{Projected sensitivity of future colliders}
\label{sec:future}
In this section, we investigate the sensitivity on the MMO at the future $e^+ e^-$ colliders such as ILC~\cite{ILC:2013jhg} and FCC-ee~\cite{Agapov:2022bhm}. Our study will remain same for the case of Muon Collider~\cite{MuonCollider:2022xlm}.  We will mainly focus on lepton collider as this will have a lot cleaner background than the hadron collider; in particular, there are hardly any pile-up events~\cite{An:2018dwb,Boscolo:2019awb}. As a result, it may be far more simpler to distinguish a hypothetical SM background from the signal coming from the RHN decay.  For illustration, we consider collision energy $\sqrt{s}=91~{\rm GeV}$ and $1~{\rm TeV}$ with an integrated luminosity of $\mathcal{L}= 150\times 10^{3}~\text{fb}^{-1}$ and $10^{3}~\text{fb}^{-1}$, respectively. We generate the UFO model files using FeynRules~\cite{Christensen:2008py} and use that in MadGraph~\cite{Alwall:2014hca} to calculate cross section and simulate signal/background events. As most of our signal consists of photon or lepton we keep our analysis up to parton level.
\subsubsection{$e^+ e^- \to n\gamma + \text{Inv}$ with $n=1,2,3$}
\begin{figure}
	\centering
\begin{subfigure}[b]{0.45\textwidth}
         \includegraphics[width=\textwidth]{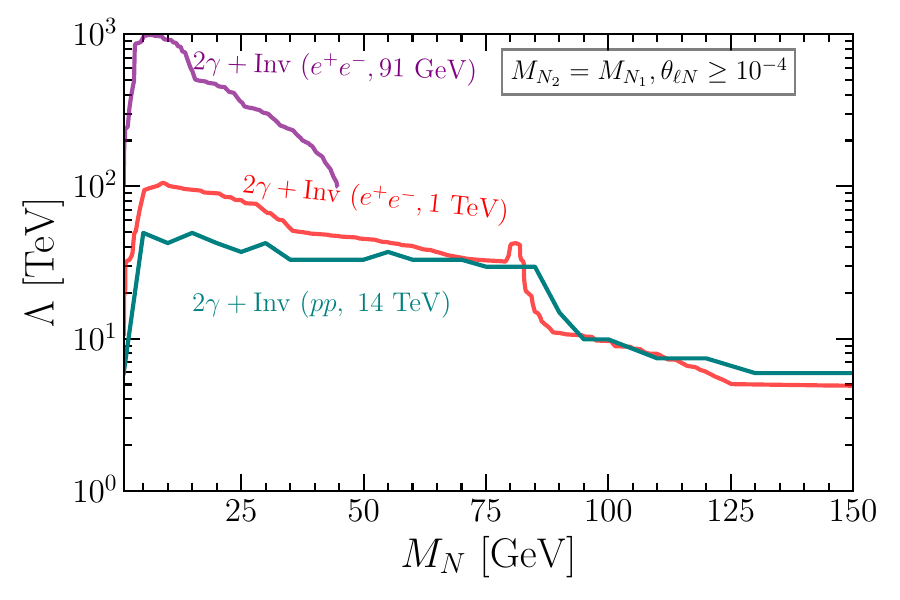}
         \caption{ }
         \label{fig:exdi}
     \end{subfigure}
     \begin{subfigure}[b]{0.45\textwidth}
         \includegraphics[width=\textwidth]{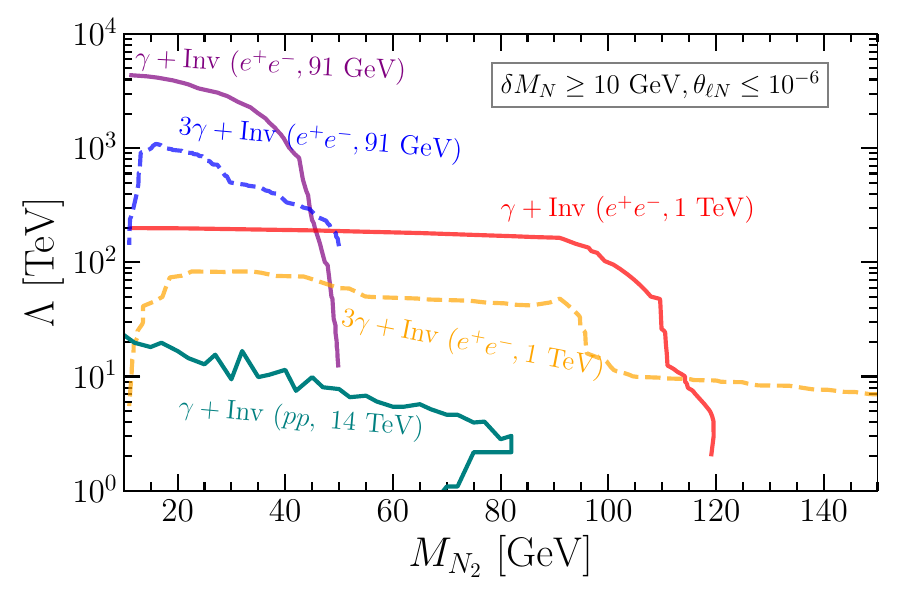}
         \caption{ }
         \label{fig:exmono}
     \end{subfigure}
      \begin{subfigure}[b]{0.45\textwidth}
         \centering
         \includegraphics[width=\textwidth]{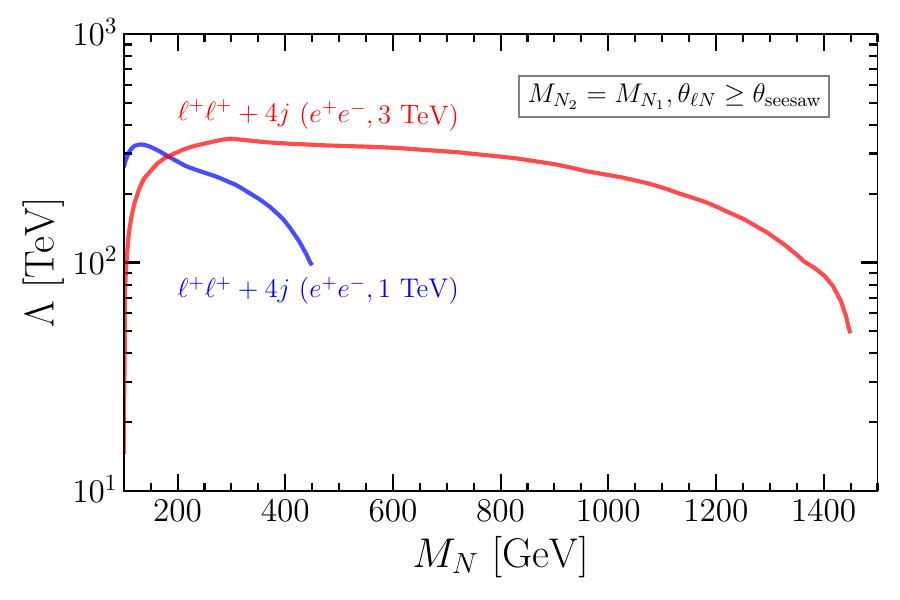}
         \caption{ }
         \label{fig:exssd}
     \end{subfigure}
 \begin{subfigure}[b]{0.45\textwidth}
         \centering
         \includegraphics[width=\textwidth]{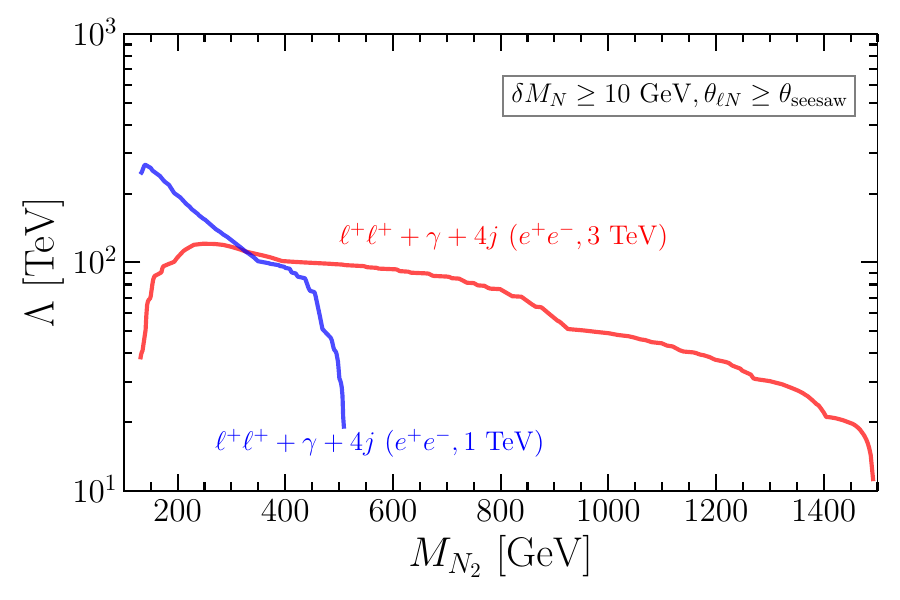}
         \caption{ }
         \label{fig:exssdgamma}
     \end{subfigure}
\caption{The expected $95\%$ C.L. projected limit in $M_N-\Lambda$ plane coming from various proposed signatures at ILC, FCC-ee or HL-LHC. The regions below each lines are excluded. The upper and lower left panels stand for degenerate~($M_{N_2}\approx M_{N_1}$) case, whereas the upper and lower right panels stand for non-degenerate~($\delta M_N>10$ GeV) case. The upper and lower panel stands for the final states $n\gamma+\text{Inv}$ and $\ell^+\ell^++4j(\gamma)$, respectively. See text for more detail.}
\label{fig:projection}
\end{figure}

We find that the dominant SM backgrounds~(BKGs) for the signal $e^+ e^- \to 2\gamma + \text{Inv}$ are $e^+ e^- \to 2\gamma + \nu\bar{\nu}$ and $e^+ e^- \to 2\gamma $. Although the $e^+ e^- \to 2\gamma $ process has large cross-section but can be greatly reduced with relatively large missing energy cut. We consider two c.m. energy $\sqrt{s}=M_Z$ and 1 TeV and the corresponding selection cuts to optimize the signal/BKGs ratios are listed in Table.~\ref{tab:future-sensitivity}. In addition to the missing energy cut, relatively large cut for transverse momenta of photons~($p_T^{\gamma_i}$)  and invariant mass of photon pair~($M(\gamma_1,\gamma_2)$) helps to reduce the BKGs further as for BKGs, $p_T^{\gamma_i}$ and $M(\gamma_1,\gamma_2)$ peaks at lower values. Note that this signal is only valid when $M_{N_2}=M_{N_1}$ and when mixing is relatively large $|\theta_{\ell N_i}|\geq 10^{-4}$ such that photons decay inside the detector. In Fig.~\ref{fig:exdi}, we show the projected sensitivity coming from this final state for $\sqrt{s}=M_Z$ and 1 TeV by purple and red lines, respectively. We see that in the case of c.m. energy $\sqrt{s}=M_Z$, this final state can probe the cut-off scale as $\Lambda\sim 10^3$ TeV depending on the heavy neutrino mass.
\par For the final state $e^+ e^-\to\gamma+\text{Inv}$, we find the dominant SM BKG is $e^+ e^-\to\gamma+\nu\bar{\nu}$. Again the missing energy cut plays important role to reduce the BKG. In Table.~\ref{tab:future-sensitivity}, we show the selection cuts to optimize the signal/BKGs ratios for our two considered c.m. energy $\sqrt{s}=M_Z$ and 1 TeV. Note that for mono-photon signal, BKG $p_T^\gamma$ distribution peaks at lower value compare to the signal and the distribution is flat as long as $M_{N_1}$ is small. As $M_{N_1}\to M_Z/2$, photon becomes softer and the high $p_T^\gamma$ cut will not be efficient as the signal cross-section also falls off. For $\sqrt{s}=1$ TeV, in addition to the basic acceptance cut as mentioned in Table.~\ref{tab:future-sensitivity}, we also utilize the recoil mass of photon defined as $M_{\rm Recoil}=\sqrt{s-2\sqrt{s} E_\gamma}$, $E_\gamma$ being the energy of photon. The BKG events shows a sharp peak at $Z$-resonance in the $M_{\rm Recoil}$ distribution, which can be removed by the following cut $M_{\rm Recoil}>120$ GeV. Note that for $\sqrt{s}=M_Z$, $M_{\rm Recoil}$ cut is not very useful as large number of signal events populates near the $Z$ peak. This signal is valid only if mass splitting is relatively large~($\delta M_N>10$ GeV) and mixing is relatively small~($\theta_{\ell N_i} < 10^{-6}$), such that $N_2$ decays promptly but $N_1$ decays outside the detector. The resulting sensitivity coming from this final state are shown in Fig.~\ref{fig:exmono} by the purple line  (red line) for $\sqrt{s}=M_Z$ ($\sqrt{s}=1$ TeV). This final state can probe the cut-off scale as $\Lambda\sim 3\times 10^3$ TeV and $\Lambda\sim 2\times 10^2$ TeV for $\sqrt{s}=M_Z$ and 1 TeV, respectively.
\par For the signal $e^+e^-\to 3\gamma+\text{Inv}$, the dominant SM BKG comes from the process $e^+e^-\to 3\gamma+\nu\overline{\nu}$ and all the relevant cuts for c.m. energy $\sqrt{s}=M_Z$ and 1 TeV to optimize signal/BKG ratio are mentioned in Table.~\ref{tab:future-sensitivity}. Here we relax the $p_T^{\gamma_3}$ cut as this photon is mostly coming from the $N_2\to N_1\gamma$ decay and is soft for small mass splitting $\delta M_N\sim 10$ GeV. We also find that one needs to relax the $\Delta R$ cuts as there are three photons for signal which are not widely separated in the $\eta-\phi$ plane. Again, the region of validity for this signal is same as of $\gamma+\text{Inv}$ and hence the results are shown in the same panel, Fig.~\ref{fig:exmono} by the blue-dashed line  (orange-dashed line) for $\sqrt{s}=M_Z$ ($\sqrt{s}=1$ TeV). We find that the resulting sensitivity is always weaker compared to $\gamma+\text{Inv}$ final state.
\small
\begin{table}[]
\centering
\begin{tabular}{|c|c|c|c|}\hline
Signal & Dominant BKGs & $\sqrt{s}$  &  Selection Cuts \\
\hline
\hline
\multirow{2}{*}{$2\gamma+\text{Inv}$} & \multirow{2}{*}{$2\gamma$, $2\gamma+\nu\bar{\nu}$} & $M_Z$ & $N^\gamma=2$, $p_T^{\gamma_1}>15$ GeV,$p_T^{\gamma_2}>10$ GeV, $\Delta R(\gamma,\gamma)>0.4$, $|\eta^\gamma|<2.44$,  \\
           &                         &                                     &   MET $>$ 10 GeV, $M(\gamma_1,\gamma_2)>40$ GeV\\ \cline{3-4}
   &  &  1 TeV & $N^\gamma=2$, $p_T^{\gamma_1}>150$ GeV, $p_T^{\gamma_2}>80$ GeV, $\Delta R(\gamma,\gamma)>0.4$,  \\ 
              &                         &                                     &   $|\eta^\gamma|<2.44$, MET $>$ 50 GeV, $M(\gamma_1,\gamma_2)>200$ GeV\\     \hline
\multirow{2}{*}{$\gamma+\text{Inv}$} & \multirow{2}{*}{$\gamma+\nu\bar{\nu}$} & $M_Z$ &  $p_T^{\gamma}>15$ GeV, $|\eta^\gamma|<2.44$ \\ \cline{3-4}   
                                     &                                        & 1 TeV &  $p_T^{\gamma}>40$ GeV, $|\eta^\gamma|<2.44$, $M_{\rm Recoil} > 120$~GeV \\ \hline 
\multirow{2}{*}{$3\gamma+\text{Inv}$} & \multirow{2}{*}{$3\gamma+\nu\bar{\nu}$} & $M_Z$ & $N^\gamma=3$, $p_T^{\gamma_1}>15$ GeV,  $p_T^{\gamma_2}>10$ GeV, $p_T^{\gamma_3}>5$ GeV, $|\eta^\gamma|<2.44$,  \\ 
                                    &                                           &       &  $\Delta R(\gamma,\gamma)>0.2$, MET $>$ 10 GeV  \\  \cline{3-4}    
                                    &                                           &  1 TeV &  $N^\gamma=3$, $p_T^{\gamma_1}>120$ GeV,  $p_T^{\gamma_2}>60$ GeV, $p_T^{\gamma_3}>15$ GeV, $|\eta^\gamma|<2.44$,  \\     
                                  &                                           &       &  $\Delta R(\gamma,\gamma)>0.2$, MET $>$ 50 GeV  \\  \hline
\end{tabular}
\caption{Signal, dominant SM BKGs and the corresponding selection cuts we used to reduce the BKGs.}
\label{tab:future-sensitivity}
\end{table}
\normalsize
\par It is straightforward to estimate the sensitivity of the HL-LHC for the final states such as $\gamma+\text{Inv}$ and $2\gamma+\text{Inv}$. The number of produced RHNs is $\mathcal{N}_{N}^{\rm prod}\propto \mathcal{L}|\alpha_{\rm NB}^{12}|^2/\Lambda^2$ where $\mathcal{L}$ is the luminosity. Assuming the selection efficiency~($\epsilon_{\rm cut}$) roughly remains same in HL-LHC and LHC, scaling behaviour of the number of events with the model parameters is also same. Also with good approximation, background scales with $\mathcal{L}$. Hence, the exclusion region of the sensitivity in the plane $M_N-\Lambda$ changes as $\mathcal{L}^{1/4}$. For HL-LHC, assuming $\mathcal{L}_{\rm HL-LHC}=3000\,\text{fb}^{-1}$, exclusion region will broaden by the factor of $(3000/39)^{1/4}\approx 3$ than it was derived in Fig.~\ref{fig:BoundLEP} for LHC. We show the projected bounds for $2\gamma+\text{Inv}$ and $\gamma+\text{Inv}$ in Fig.~\ref{fig:exdi} and \ref{fig:exmono}, respectively by dark cyan line. 
\subsubsection{$ e^+ e^- \to \ell^\pm \ell^\pm + 4j/ \ell^\pm \ell^\pm + 4j+\gamma$}
Now let us discuss the projected sensitivity coming from the signal processes such as $ e^+ e^- \to \ell^\pm \ell^\pm + 4j/ \ell^\pm \ell^\pm + 4j+\gamma$. For these type of signals we choose higher c.m. energy such as $\sqrt{s}=1$ and 3 TeV as these are only valid when $M_{N}>100$ GeV.  The dominant SM BKGs for the signal $ e^+ e^- \to \ell^\pm \ell^\pm + 4j$ comes from the SM processes such as $ e^+ e^- \to W^\pm W^\pm W^\mp W^\mp (W^\pm W^\pm W^\mp jj) \to \ell^\pm \ell^\pm +4j+\text{MET}$. At $\sqrt{s}=1(3)$ TeV, the SM background has cross section $\sigma=1.2(2.8) \times 10^{-5}$ pb after basic acceptance cuts:  $p_T^{\ell,j}\ge30$ GeV, $|\eta^{\ell}|<2.44$, $|\eta^{j}|<5$, and  $\Delta R(\ell,\ell)>0.4$. In addition we demand $\Delta R(\ell,j)>0.2$ during signal selection as given below. This is to ensure leptons are isolated from the hadronic activity. The signal selection cuts is defined as
\begin{equation}	
N^\ell=2,~|\sum\ell_{\text{charge}}|=2,~\Delta R(\ell,j)>0.2,~ \text{MET}\le 10~(20)~\text{GeV},~p_T^{\ell_1}\ge 80~(200)~\text{GeV},~p_T^{\ell_2}\ge 30~(60)~\text{GeV}
\label{eq:ssd3tev}
\end{equation}
With the above selection cuts, we obtain $99\%$ background reduction and $12\%-56\%$~($36\%-82\%$) of signal events are selected for $M_N$ in range $100-500$~GeV ($200-1500$~ GeV). For $M_N=100$ GeV, the leptons are not isolated from the jets so the cuts $\Delta R(\ell,j)>0.2$ kills almost all signal events. This signal is valid as long as $M_{N_2}\approx M_{N_1}$ and $M_{N_i}>100$ GeV. The signal is true for any value of light-heavy neutrino mixing as long as $N_i$ decays promptly, which we find to be always true for mixing $\theta_{\ell N_i}>\theta_{\rm seesaw}$. The resulting sensitivity are shown in Fig.~\ref{fig:exssd} for c.m. energy $\sqrt{s}=1$ TeV and 3 TeV with blue and red lines, respectively. We see that this can probe the cut-off scale as $\Lambda\sim 300$ TeV at ILC or FCC-ee.
\par The SM background for the signal $ e^+ e^- \to \ell^\pm \ell^\pm + 4j+\gamma$ can arise from the following process: $e^+ e^- \to  W^\pm W^\pm W^\mp W^\mp +\gamma \to \ell^\pm \ell^\pm +4j+\gamma+\text{MET}$. At $\sqrt{s}=1$ TeV, the cross section for $e^+ e^- \to  W^\pm W^\pm W^\mp W^\mp +\gamma $ is $\sigma=4 \times 10^{-5}$ pb. After taking into account the branching fraction for of $W^\pm$ boson to respective final state the cross-section reduce to  $\sigma=1.9 \times 10^{-6}$ pb, which will further reduce after applying the detector acceptance cuts. We obtain $\sigma=6 \times 10^{-7}$ pb after basic acceptance cuts:  $p_T^{\ell,j}\ge30$ GeV, $p_T^{\gamma}\ge20$ GeV, $|\eta^{\ell,\gamma}|<2.44$, $|\eta^{j}|<5$, $\Delta R(\ell,\ell)>0.4$, $\Delta R(\ell,\gamma)>0.2$ and $\Delta R(\ell,j)>0.2$. As the background cross section is negligible we do not use additional signal selection cuts. The same set of basic acceptance cuts is applied for $\sqrt{s}=3$ TeV and with the similar argument as for the case of  $\sqrt{s}=1$ TeV we do not impose extra signal selection cuts. The signal valid for the parameter space $M_{N_1}>100$ GeV, $\delta M_N>10$ GeV and $\theta_{\ell N_i}>\theta_{\rm seesaw}$. In Fig.(~\ref{fig:exssdgamma}), we show the $95\%~$CL exclusion limit in $M_{N_2}-\Lambda$ plane by the blue~($\sqrt{s}=1$ TeV) and red line~($\sqrt{s}=3$ TeV) coming from  the final state $\ell^\pm \ell^\pm + 4j+\gamma$. We find that although the sensitivity coming from this final state is weaker compare to $\ell^\pm \ell^\pm + 4j$, but might be improved with the assumption of larger mass splitting $\delta M_N$.
\section{Bounds from astrophysics and cosmology}
\label{sec:astro}
In addition to the existing collider constraints, the MMO is subject to various astrophysical constraints from stellar objects such as Red giants and supernova as well as from Big Bang Nucleosynthesis~(BBN)~\cite{Brdar:2020quo,Vogel:2013raa,Diaz:2019kim,Magill:2018jla,Boyarsky:2020dzc,Bondarenko:2021cpc,Sabti:2020yrt,Plestid:2020vqf}. Although these type of constraints are relevant only for light mass of heavy neutrinos but for the sake of completeness we discuss them in the following.\\
\underline{\bf{Stellar cooling:}} The cooling of red giant stars plays a prominent role to constrain the MMO. In presence of this operator, the dispersion relation of electromagnetic excitations~(referred to as plasmons) is altered inside the hot plasma that forms the core of a star. The plasmons acquire a temperature dependent mass and hence in presence of MMO, the decay channel for the plasmons into a neutrino pair will open up. If produced these neutrinos will leave the star unhindered, resulting in an additional cooling mechanism. As a result this can be used to impose stringent upper limit on the MMO.
\par
\begin{figure}[!htbp]
	\centering
	\includegraphics[width=0.65\linewidth]{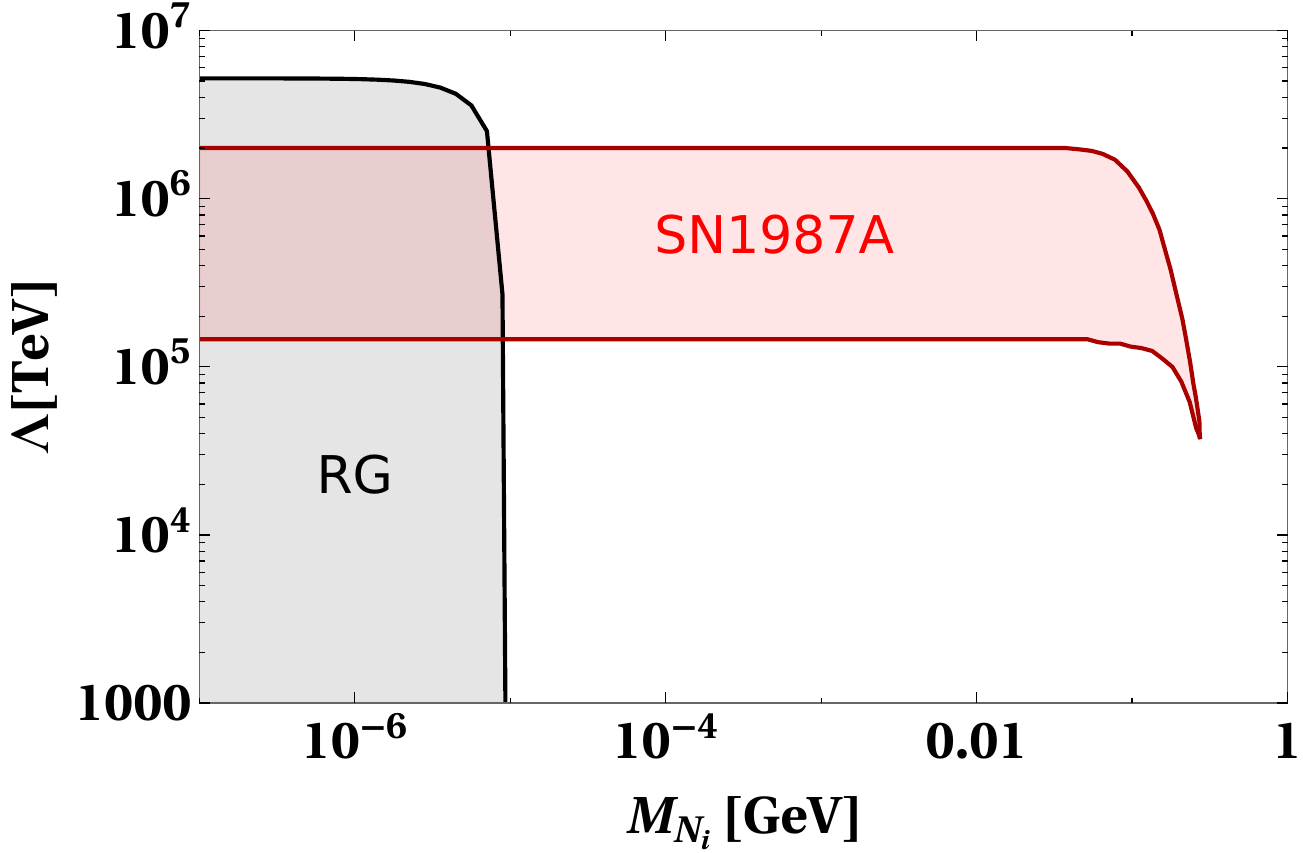}
	\caption{Disallowed region in the $M_N-\Lambda$ plane from red-giant~\cite{Vogel:2013raa,Brdar:2020quo,Raffelt:1996wa,Diaz:2019kim} and supernova cooling~\cite{Magill:2018jla,Kadota:2014mea} by black and red shaded regions, respectively. If the cut-off scale is $\Lambda\gtrsim 2\times 10^{6}$ TeV, the production of RHNs is not enough to efficiently cool the interior of supernova, whereas for the cut-off scale $\Lambda\lesssim 1.4\times 10^5$ TeV, RHNs interact too frequently and are trapped inside the supernova.}
	\label{fig:BoundStar}
\end{figure}

 Let us now calculate the decay width of these plasmons to heavy neutrino pair $N_1 N_2$ which is given as~\footnote{The same operator can also induce plasmons decay to $N-\nu$ but the relevant couplings for this process are suppressed by mixing $\theta$ and hence did not consider in our analysis.}
\begin{align}
\Gamma_{\gamma^*}=\frac{|\mu_{NN}|^2}{24\pi}\frac{K^4}{\omega},\,\,\text{with}\,\,
|\mu_{NN}|^2=\frac{16 c_W^2 |\alpha_{NB}^{12}|^2}{\Lambda^2} f_Z\Big(1,\frac{M_{N_1}}{K},\frac{M_{N_2}}{K}\Big)\lambda^{\frac{1}{2}}\Big(1,\frac{M_{N_1}^2}{K^2},\frac{M_{N_2}^2}{K^2}\Big).
\end{align}
In the above $\omega$ and $k$ are the plasmon energy and momentum, respectively. $K=\sqrt{\omega^2-k^2}$ is the effective plasmon mass. Note that in a non-relativistic non-degenerate plasma the emissivity of neutrinos is dominated by transverse plasmons~\cite{Brdar:2020quo} which have an effective mass~($K$) equal to the plasma frequency $\omega_p$. The energy loss per unit volume of transverse plasmon decay into $N_1 N_2$ is~\cite{Vogel:2013raa}
\begin{align}
Q=\int_0^{\infty}\frac{k^2 dk}{\pi^2}\,\frac{\omega\Gamma_{\gamma^*\to N_1 N_2}}{e^{(\omega/T_\gamma)}-1}\Big|_{\omega=\sqrt{\omega_p^2+k^2}},
\label{eq:loss}
\end{align}
where $T_\gamma$ is the plasma temperature. We use the recent analysis of global clusters from Ref.~\cite{Raffelt:1996wa}, which sets an upper limit on the active neutrino magnetic moment $|\mu_\nu|<2.2\times 10^{-12}\,\mu_B$ using the plasma characteristic for a red-giant core with $\omega_p=18$ KeV and $T_\gamma=8.6$ KeV~\cite{Diaz:2019kim}. We recast this bound for the massive heavy neutrino case by equating the energy loss~(Eq.~\ref{eq:loss}) for massive and massless $N$, and solve for the $\mu_{NN}$. The obtained limit on $\mu_{NN}$ is shown as a function of $M_{N}$ and $\Lambda$ in the left panel of Fig.~\ref{fig:BoundStar} as the gray exclusion region. Note that for this we assume degenerate heavy neutrino masses.
\par Although the bound coming from red-giant cooling can be tight, it can only be applied to RHNs that are relatively light. Hence to extend the exclusion limit for higher RHN mass one should consider denser, hotter astrophysical objects as the corresponding plasma frequency $\omega_p$ is larger. Unfortunately, the limits obtained on $\mu_{NN}$ from these objects are so poor that the exclusion limit on $\Lambda$ is not competitive with bounds coming from red-giant star.\\
\underline{\bf{Supernova 1987A:}} The same neutrino electromagnetic coupling which is responsible for stellar cooling also can give a new supernova cooling mechanism through the process $e^+e^-\to N_1 N_2$, with $N$ escaping~\footnote{There will be additional contribution from the process $\gamma+\nu\to N_i$ but are suppressed as the amplitude is proportional to $\theta\alpha_{NB}/\Lambda$, and hence neglected for simplicity.}. This will open up an efficient energy sink, which implies faster cooling of the proto-neutron star~\cite{Raffelt:1996wa,Magill:2018jla}. We expect that in the case of degenerate mass~($\delta M_N=0$) our scenario will be qualitatively comparable to the one examined in Ref.~\cite{Kadota:2014mea}. In Fig.~\ref{fig:BoundStar} we show the adopted corresponding exclusion region to our case. We find that in the mass range $M_{N_i}\lesssim 100$ MeV, these limits exclude the cut-off scale in the range $1.4\times 10^{5}\text{ TeV}\lesssim \Lambda \lesssim 2\times 10^{6}\text{ TeV}$. The constraint disappear if $\Lambda$ lies outside this region. If the cut-off scale $\Lambda$ is too large, the number of RHNs produced will be very few and hence not enough to efficiently cool the interior of the supernova. On the other hand, if the cut-off scale $\Lambda$ is small enough then RHN interact too frequently to leave the supernova core and hence do not contribute to the cooling of the core. Note that the sensitivity on the cut-off scale $\Lambda$ is almost flat for low mass as the emission rate is not so sensitive when $M_{N_i}\ll T$, where $T\sim 30$ MeV is the supernova core temperature.  The constraint becomes weaker when $M_{N_i}>100$~MeV as the RHNs are too heavy to be produced in plasmon decays.
\par For the case of $\delta M_N\neq 0$, we expect the supernova limit to be quite different from the degenerate case, but given the difficulty of performing such analysis, we leave this for future work. \\
\underline{\bf{BBN:}} Cosmology can be very useful for constraining the coupling of metastable particles. Their presence in the early universe implies modified expansion rate and make $p\leftrightarrow n$ interactions freeze out faster, leading to a larger neutron-proton ratio. In conclusion this can spoil the predictions of the standard BBN model. This constraints the lifetime of heavy neutrinos to be $\tau_N\lesssim 10^{-2}$ sec~\cite{Boyarsky:2020dzc,Bondarenko:2021cpc,Sabti:2020yrt}. In the non-degenerate case, heavier neutrino $N_2$ decay width is controlled mainly by the dimension five operator and as long as mass splitting is large, $N_2$ decays promptly. For example with $\delta M_N=10^{-3}$ GeV, 1 GeV and 10 GeV, $N_2$ decays fast enough to avoid BBN bounds if the cut-off scale is $\Lambda\lesssim 10^{6}$ GeV, $10^{11}$ GeV and $10^{12}$ GeV, respectively. On the other hand lightest of heavy neutrinos $N_1$ decay width or lifetime is determined by its mixing with active neutrinos and is already shown in~\cite{Boyarsky:2020dzc} that BBN exclude $M_{N_1}\lesssim (0.4-0.5)$ GeV, if $N_1$ mixes dominantly with $\nu_e$ or $\nu_\mu$, while lighter masses can be allowed for mixing dominantly with $\nu_\tau$. The situation becomes less constrained considering three RHNs as one can easily allow large $N_1-\nu_\tau$ mixing.\\
\underline{\bf{Dark matter:}} In addition to the above mentioned constraints, further constraint exists if one of the heavy neutrinos plays the role of dark matter through the MMO~\cite{Cho:2021yxk}. We know that neutrino oscillation data can be explained only with 2 RHNs. Hence if we add 3rd RHN and if it does not couple with light neutrinos, it can play the role of dark matter candidate. The dominant decay mode for $N_3$ is $N_3 \to \nu\gamma$ and is suppressed as it is proportional to the product of off-diagonal MMO and mixing of $N_{1,2}$ with light neutrinos,
\begin{align}
\Gamma(N_3 \rightarrow \nu \gamma)
\sim  \frac{1}{10^{28} \text{ sec}} \left(\frac{10^{15} \text{ GeV}}{\Lambda_5}\right)^2 \left(\frac{|\Theta|}{10^{-6}}\right)^2 \left(\frac{M_{N_3}}{1 \text{ MeV}}\right)^3,
\end{align}
where $\Theta=\sum_{k=1}^2\sum_{j=1}^3\theta_{jk}(\alpha_{NB})_{k3}$. Hence the life time can be even longer than $10^{28}$ sec which is required to be a dark matter candidate for $|\Theta|=10^{-6}$, $M_{N_3}\sim \mathcal{O}(\text{MeV})$ and $\Lambda \gtrsim 10^{15}$ GeV. Relic density can also be explained through the thermal and non-thermal production of $N_3$ via MMO with the mass around MeV scale and $\Lambda \gtrsim 10^{15}$ GeV. For a detailed discussion on this, see Ref.~\cite{Cho:2021yxk}.
\par From the above discussions we can draw the following conclusions: There are tight bounds on the cut-off scale $\Lambda \gtrsim 4\times 10^{6}$ TeV coming from red giants cooling for the mass range $M_N\lesssim 10^{-5}$ GeV; For mass range $M_N \lesssim 0.1$ GeV supernova cooling provides bound on the cut-off scale in the range $1.4\times 10^{5}\text{ TeV}\lesssim \Lambda \lesssim 2\times 10^{6}\text{ TeV}$; BBN on the other hand does not put any meaningful constraint as long as mass splitting is relatively large; If one of the heavy neutrinos plays the role of dark matter then for MeV scale mass, the cut-off scale should be $\Lambda\gtrsim 10^{15}$ GeV. 
\par Note that this new operaor contains new sources of CP non-conservation and hence potentially can modify the standard leptogenesis scenarios which in turn can give bound on the cut-off scale $\Lambda$. More specifically, it can contributes to the relevant lepton number violating decays $N\to\ell^-\phi^+$. We find that this has been somewhat discussed in Ref.~\cite{Aparici:2009fh} but dedicated analyses are missing in the literature.
\section{Concluding Remarks}
\label{sec:conclusion}
We analyzed novel phenomena of heavy Majorana neutrinos via dimension five operators, specifically, through the magnetic MMO. We find that unlike the simplest type-I seesaw, MMO offers new unsuppressed heavy neutrino production mechanism at various colliders such as $e^+e^-$ or $pp$. The heavy neutrinos can be produced at $e^+e^-/pp$ collider through MMO as $e^+e^-/pp\to N_1 N_2$. These production rates are proportional to $(\alpha_{NB}/\Lambda)^2$ and hence can be sizeable even for relatively large cut-off scale $\Lambda$, see Fig.~\ref{fig:eeXS} and \ref{fig:ppXS}. Note that in type-I seesaw, these type of production rates are suppressed as these are proportional to $\theta^4$. In addition to the usual heavy neutrino decay modes coming purely from the mixing $\theta$, there are additional decay modes such as $N_i\to N_j\gamma/N_j Z/\nu_j\gamma$ due to the presence of MMO.  There appear many interesting collider signatures in the final states from the production process $e^+e^-/pp\to N_1 N_2$ followed by the decay chain such as $N_i\to N_j\gamma/N_j Z/\nu_j\gamma$ and $N_i\to\ell^\pm jj$. Among them, the most interesting collider signatures we find are  $n\gamma + \text{missing}$ ($n=1,2,3$), and $\ell^\pm \ell^\pm +4j (+\gamma)$. Identifying the parameter regions already constrained by LEP and LHC data, we discussed the discovery prospect at future colliders. In addition, we also discussed the various astrophysical constraints from stellar objects such as Red giants and supernova as well as from BBN which are applicable for the heavy mass of $\mathcal{O}(100\,\text{MeV})$ or below. We stress that the signatures such as $\ell^\pm \ell^\pm +4j (+\gamma)$ are lepton number violating and the cross-section can be large. Their possible detection could provide an indirect test of the Majorana nature of neutrinos, complementary to that provided by neutrinoless double beta decay searches. Note that the discussion presented in our paper can easily be extended to the proposed muon collider~\cite{Delahaye:2019omf,Li:2023tbx}. In this case one also expects a plethora of interesting collider signatures arising from the dimension five magnetic moment operator. We leave this for future work.
\section*{Acknowledgements}
The Authors thank Martin Hirsch for discussions and useful comments. The work of S.M. is supported by KIAS Individual Grants (PG086001) at Korea Institute for Advanced Study.
\appendix
\section{Decay Widths}
\label{app:decay-widths}
Here we discuss the various possible decay modes of the heavy neutrinos $N_i$ in presence of dimension five operators. In dimension four, the relevant vertices are $W-\ell-N$, $Z-\nu-N$ and $h-\nu-N$. In dimension five, there will be additional vertices such as $Z/\gamma-N-N$, $Z/\gamma-\nu-N$ and $h-\nu-N$, $h-N-N$ from operators $\overline{N_R^c}\sigma_{\mu\nu}N_R B^{\mu\nu}$ and $\overline{N_R^c}N_R (H^\dagger H)$, respectively. The analytical expressions for different two and three body partial decay widths of the RH neutrinos $N_i$ are given as:
\subsection{Two body decay widths}
\begin{align}
&\Gamma(N_i\to\ell_j^- W^+) = |\theta_{ji}|^2 \frac{G_f}{8\sqrt{2}\pi M_{N_i}^3} (M_{N_i}^2 - M_W^2) \left( M_{N_i}^2 (M_{N_i}^2 + M_W^2) - 2M_W^4 \right), \\
&\Gamma(N_i\to\nu_j Z) = |\theta_{ji}|^2 \frac{G_f}{8\sqrt{2}\pi M_{N_i}^3} (M_{N_i}^2 - M_Z^2) \left( M_{N_i}^2 (M_{N_i}^2 + M_Z^2) - 2M_Z^4 \right), \\
& \Gamma(N_i\to \nu_j h) = \frac{|\theta_{ji}|^2}{16\pi M_{N_i} v^2} \left( M_{N_i}^2 - m_h^2 \right), \\
& \Gamma(N_i\to\nu_j\gamma) = \frac{2 c_W^4 M_{N_i}^3}{\pi \Lambda^2} \Big|\sum_{k=1}^2 \theta_{jk} |(\alpha_{NB})_{ki}\Big|^2,\\
& \Gamma(N_i\to N_j\gamma) = \frac{2 c_W^2 |\alpha_{NB}^{ji}|^2}{\pi \Lambda^2 M_{N_i}^3} (M_{N_i}^2 - M_{N_j}^2 )^3,\\
& \Gamma(N_i\to N_j h) = \frac{v^2 |\alpha_{NH}^{ji}|^2}{4\pi \Lambda^2 M_{N_i}} \left( (M_{N_i}^2 + M_{N_j}^2 - m_h^2) + 2 M_{N_i} M_{N_j} \cos (2\delta_{NH}^{ji})\right) \lambda^{\frac{1}{2}}\left(1,\frac{m_h^2}{M_{N_i}^2}, \frac{M_{N_j}^2}{M_{N_i}^2}\right), \\
& \Gamma(N_i\to N_j Z) = -\frac{s_W^2 |\alpha_{NB}^{ji}|^2}{\pi \Lambda^2 M_{N_i}} f_{Z}\left(M_Z, M_{N_i}, M_{N_j}\right)\, \lambda^{\frac{1}{2}}\left(1,\frac{M_{N_j}^2}{M_{N_i}^2}, \frac{M_Z^2}{M_{N_i}^2}\right),
\end{align}
where $\lambda(x,y,z)=x^2+y^2+z^2-2xy-2xz-2yz$, $\alpha_{NH}^{ji}=|\alpha_{NH}^{ji}| e^{i\delta_{NH}^{ji}}$, $\alpha_{NB}^{ji}=|\alpha_{NB}^{ji}| e^{i\delta_{NB}^{ji}}$ and
\begin{align}
 f_{Z}(M_Z,M_{N_i},M_{N_j})=\left( M_Z^2 (M_{N_i}^2 + M_{N_j}^2) - 2(M_{N_i}^2-M_{N_j}^2)^2 + M_Z^4 - 6 M_{N_i} M_{N_j} M_Z^2\cos (2\delta_{NB}^{ji})\right)
\end{align}
\subsection{Three body decay widths}
\begin{align}
& \Gamma(N_i\to\ell_j^-\ell_k^+\nu_k; j\ne k)  =|\theta_{ji}|^2 \frac{G_f^2 M_{N_i}^5}{64\pi^3} I_1(x_{\nu_k},x_{\ell_k},x_{\ell_j}),\\
&\Gamma(N_i\to \nu_j\ell_j^-\ell_j^+)  =|\theta_{ji}|^2 \frac{G_f^2 M_{N_i}^5}{64\pi^3} \Big\{ \left((g_L^\ell)^2 + 2(g_R^\ell)^2 + (g_L^\ell - 1)^2\right) I_1(x_{\ell_j},x_{\ell_j},x_{\nu_j}) \nonumber \\
& + 8 g_R^\ell (2g_L^\ell-1) I_2(x_{\ell_j},x_{\ell_j},x_{\nu_j}) \Big\}, \\
&\Gamma(N_i\to \nu_j\ell_k^-\ell_k^+; j\ne k)  =|\theta_{ji}|^2 \frac{G_f^2 M_{N_i}^5}{32\pi^3} \Big\{ \left((g_L^\ell)^2 + (g_R^\ell)^2 \right) I_1(x_{\ell_k},x_{\ell_k},x_{\nu_j}) \nonumber \\
& + 8 g_R^\ell g_L^\ell I_2(x_{\ell_k},x_{\ell_k},x_{\nu_j}) \Big\}, \\
&\Gamma(N_i\to\nu_j\nu\nu)=|\theta_{ji}|^2 \frac{G_f^2 M_{N_i}^5}{512\pi^3}, \\
& \Gamma(N_i\to \ell_j^- u_\alpha \overline{d_\beta} ) = N_c |\theta_{ji}|^2 |V_{\rm CKM}^{\alpha\beta}|^2 \frac{G_f^2 M_{N_i}^5}{64\pi^3} I_1(x_{u_\alpha},x_{d_\beta},x_{\ell_j}), \\
& \Gamma(N_i\to \nu_j q\bar{q})= N_c |\theta_{ji}|^2 \frac{G_f^2 M_{N_i}^5}{32\pi^3} \Big\{ \left( (g_L^q)^2 + (g_R^q)^2 \right) I_1(x_{q},x_{q},x_{\nu_j}) + 8 g_R^q g_L^q I_2(x_{q},x_{q},x_{\nu_j}) \Big\},
\end{align}
where,
\begin{align}
& I_1(x_a, x_b, x_c)=\int_{(x_a+x_b)^2}^{(1-x_c)^2} \frac{dz}{z} (z-x_a^2-x_b^2) (1+x_c^2-z)\lambda^{\frac{1}{2}}(z, x_a^2, x_b^2) \lambda^{\frac{1}{2}}(1,x_c^2,z), \\
& I_2(x_a, x_b, x_c)=x_a x_b \int_{(x_a+x_b)^2}^{(1-x_c)^2} \frac{dz}{z} (1+x_c^2-z)\lambda^{\frac{1}{2}}(z, x_a^2, x_b^2) \lambda^{\frac{1}{2}}(1,x_c^2,z),\\
& g_L^\ell=-\frac{1}{2}+s_W^2, g_R^\ell=s_W^2, g_L^u=\frac{1}{2}-\frac{2}{3} s_W^2, g_R^u=-\frac{2}{3} s_W^2, g_L^d=-\frac{1}{2}+\frac{1}{3} s_W^2, g_R^d=\frac{1}{3} s_W^2.
\end{align}
\section{Cross section}
\label{app:cross-section}
The total differential production cross-section for the process $e^+e^-\to N_1 N_2$ is calculated as
\begin{align}
\frac{d\sigma}{d\cos\theta}=\frac{8e^2}{32\pi s^2\Lambda^2} |\alpha_{NB}^{12}|^2 \left( s^2 - \cos^2\theta \,\lambda(s, M_{N_1}^2, M_{N_2}^2) - (M_{N_1}^2-M_{N_2}^2)^2 - 4 s M_{N_1} M_{N_2}\cos(2\delta_{NB}^{12})\right) F_f(s),
\label{eq:diff-distribution}
\end{align}
where,
\begin{align}
F_f(s) = \frac{|\chi(s)|^2}{c_W^2} \left( (g_L^f)^2 + (g_R^f)^2 \right) + 2 c_W^2 + 2 \text{Re} [\chi(s)] (g_L^f + g_R^f),\,\,\,  \chi(s)=\frac{s}{s-M_Z^2 + i \Gamma_Z M_Z}.
\end{align}
Performing the integration over $\cos\theta$ in the range $[-1,1]$, we get the total cross-section as,
\begin{align}
\sigma=\frac{e^2}{3\pi\Lambda^2 s^2} |\alpha_{NB}^{12}|^2 F_N(\sqrt{s},M_{N_1},M_{N_2}) F_f(s),
\end{align}
where,
$F_N(m_a,m_b,m_c)=\left( m_a^2 (m_a^2 + m_b^2 + m_c^2) - 2(m_b^2-m_c^2)^2 - 6 m_a^2 m_b m_c \cos (2\delta_{NB}^{12})\right) \lambda^{\frac{1}{2}} \left(1,\frac{m_b^2}{m_a^2},\frac{m_c^2}{m_a^2}\right).$

Similarly we can calculate the helicity cross-section for the process $e_L^- e_R^+\to N_1 N_2$ and $e_R^- e_L^+\to N_1 N_2$ as follows,
\begin{align}
& \sigma(e_L^- e_R^+\to N_1 N_2)=\frac{4e^2}{3\pi\Lambda^2 s^2} |\alpha_{NB}^{12}|^2 F_N(\sqrt{s},M_{N_1},M_{N_2}) F_f^L(s), \\
& \sigma(e_R^- e_L^+\to N_1 N_2)=\frac{4e^2}{3\pi\Lambda^2 s^2} |\alpha_{NB}^{12}|^2 F_N(\sqrt{s},M_{N_1},M_{N_2}) F_f^R(s).
\end{align}
where,
\begin{align}
F_f^L(s) = \frac{|\chi(s)|^2}{c_W^2}  (g_L^f)^2 + c_W^2 + 2 \text{Re} [\chi(s)] g_L^f,\,\,\,  F_f^R(s) = \frac{|\chi(s)|^2}{c_W^2}  (g_R^f)^2 + c_W^2 + 2 \text{Re} [\chi(s)] g_R^f.
\end{align}
\bibliographystyle{utphys}
\bibliography{bibliography}
\end{document}